\documentclass[10pt,journal,compsoc]{IEEEtran}
\usepackage{amsmath,amsfonts}
\usepackage{algorithmic}
\usepackage{graphicx}
\usepackage{textcomp}
\usepackage{xcolor}
\usepackage{bm}
\usepackage[linesnumbered,ruled]{algorithm2e} 
\usepackage{algorithmic}
\usepackage{url}
\usepackage{bbding}
\usepackage{balance}
\usepackage{booktabs}
\usepackage{makecell}
\usepackage{multirow}
\usepackage{multicol}
\usepackage{hyperref}
\usepackage{proof}
\usepackage{amsmath,amsfonts}
\usepackage{amssymb}

\usepackage{setspace}
\usepackage{amsmath}
\usepackage{booktabs}
\usepackage{pifont}
\usepackage{adjustbox}
\usepackage{makecell}
\usepackage{mathtools}
\usepackage{mathtools}
\usepackage{enumitem}
\usepackage{colortbl}
\usepackage{subfigure}
\definecolor{c1}{HTML}{81658D}
\definecolor{c2}{HTML}{2a9d8f}
\definecolor{c3}{HTML}{DC925A}
\usepackage[misc]{ifsym}

\newcommand{\revise}[1]{{\color{black} #1}}
\newcommand{\round}[1]{{\color{black} #1}}

\ifCLASSOPTIONcompsoc
  \usepackage[nocompress]{cite}
\else
  \usepackage{cite}
\fi

\ifCLASSINFOpdf
\else
\fi

\begin{document}
%
\title{Improving Sequential Recommendations via Bidirectional Temporal Data Augmentation with Pre-training}
%
%
%
%

\author{Juyong Jiang, Peiyan Zhang, Yingtao Luo, Chaozhuo Li, Jae Boum Kim, Kai Zhang, Senzhang Wang, Sunghun Kim, and Philip S. Yu, \IEEEmembership{Fellow, IEEE}

\thanks{This work was supported in part by the National Natural Science Foundation of China under Grant No. 62276099. This work was done when Juyong was AI Global Residency of Upstage. Juyong Jiang, Peiyan Zhang, and Yingtao Luo are the Equal Contribution. (Corresponding authors: Chaozhuo Li; Kai Zhang.)

Juyong Jiang and Sunghun Kim are with The Hong Kong University of Science and Technology (Guangzhou), Guangzhou, China (e-mail: jjiang472@connect.hkust-gz.edu.cn; hunkim@cse.ust.hk). 

Peiyan Zhang and Jae Boum Kim are with The Hong Kong University of Science and Technology, Hong Kong SAR, China (e-mail: pzhangao@cse.ust.hk; jbkim@cse.ust.hk). 

Yingtao Luo is with Carnegie Mellon University, Pittsburgh, USA (e-mail: yingtaoluo@cmu.edu). 

Chaozhuo Li is with Microsoft Research Asia, Beijing, China (e-mail: cli@microsoft.com). 

Kai Zhang is with East China Normal University, Shanghai, China (e-mail: kzhang@cs.ecnu.edu.cn). 

Senzhang Wang is with Central South University, Hunan, China (e-mail: szwang@csu.edu.cn). 

Philip S. Yu is with the Department of Computer Science, University of Illinois at Chicago, Chicago, USA (e-mail: psyu@uic.edu).

}
}

\markboth{}%
{J.Y. Jiang \MakeLowercase{\textit{et al.}}: Improving Sequential Recommendations via Bidirectional Temporal Data Augmentation with Pre-training}
%

\IEEEtitleabstractindextext{%
\begin{abstract}
Sequential recommendation systems are integral to discerning temporal user preferences. Yet, the task of learning from abbreviated user interaction sequences poses a notable challenge. Data augmentation has been identified as a potent strategy to enhance the informational richness of these sequences. Traditional augmentation techniques, such as item randomization, may disrupt the inherent temporal dynamics. Although recent advancements in reverse chronological pseudo-item generation have shown promise, they can introduce temporal discrepancies when assessed in a natural chronological context. In response, we introduce a sophisticated approach, Bidirectional temporal data Augmentation with pre-training (BARec). Our approach leverages bidirectional temporal augmentation and knowledge-enhanced fine-tuning to synthesize authentic pseudo-prior items that \emph{retain user preferences and capture deeper item semantic correlations}, thus boosting the model's expressive power. Our comprehensive experimental analysis on five benchmark datasets confirms the superiority of BARec across both short and elongated sequence contexts. Moreover, theoretical examination and case study offer further insight into the model's logical processes and interpretability. The source code for our study is publicly available at \textcolor{blue}{\href{https://github.com/juyongjiang/BARec}{https://github.com/juyongjiang/BARec}}.
\end{abstract}

}

\maketitle

\IEEEdisplaynontitleabstractindextext

%
\IEEEpeerreviewmaketitle

\IEEEraisesectionheading{\section{Introduction}\label{sec:introduction}}
\IEEEPARstart{R}{ecommender} systems serve as pivotal engines that drive the functionality of Business-to-Consumer commercial platforms such as \emph{YouTube} and \emph{Amazon} shopping. 
Sequential recommendation (SR), in particular, represents recommender systems that entail the encapsulation of the  
sequential ordering of user actions, which characterizes the influence of past user behaviors on both current and future actions~\cite{liu2016context,liu2021augmenting,yu2016dynamic,ma2019hierarchical}. Transformer-based models~\cite{kang2018self,jiang2023adamct,sun2019bert4rec}, with their ability to handle sequential dependencies, have consistently demonstrated remarkable performance in many SR tasks.
\begin{figure}[t]
	\centering
	\includegraphics[width=0.84\linewidth]{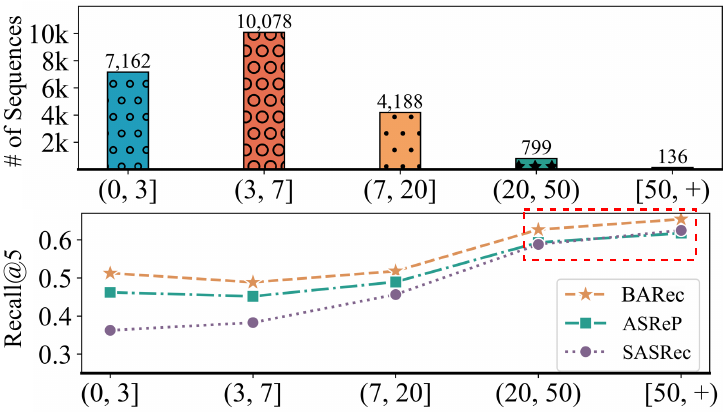}
	\caption{
 Performance (Recall@5) w.r.t sequence length distribution (bar) on \texttt{Amazon Beauty} when using (1) no sequence augmentation (SASRec, \textcolor{c1}{purple dots}); (2) reverse sequence augmentation (ASReP, \textcolor{c2}{green squares}) and (3) our bidirectional temporal augmentation (BARec, \textcolor{c3}{yellow stars}, being particularly advantageous on short sequences). In the case of long sequences with $\ge 20$, ASReP either matches or underperforms relative to SASRec, as shown in \textcolor{red}{red dotted rectangle}.
	}
	\label{fig:distribution}
\end{figure}

Despite the notable achievements attained by state-of-the-art Transformers in SR, researchers have drawn attention to their limited effectiveness when confronting short sequences that encompass fewer items \cite{liu2021augmenting}. With limited information about the user sequential patterns, prediction within this context becomes considerably more challenging \cite{wang2021sequential}.  
In addition, short sequences are prevalent or dominant in many benchmarks \cite{liu2021contrastive,liu2021augmenting,qiu2022contrastive}. 
For instance, an examination of the histogram of the Amazon Beauty dataset (Fig. \ref{fig:distribution}) reveals that a significant 32\% of sequences are very short ($L\le 3$), whilst a mere 4.2\% extend to a more substantial length ($L\ge 20$).

\round{
To solve data sparsity and short sequences problems in SR, most existing works employ a straightforward strategy to enrich the item embeddings by incorporating rich side information \cite{vasile2016meta,chen2018collective,liu2021noninvasive} or content features \cite{deldjoo2020recommender,kulkarni2020context}, providing a more contextual and holistic view of items within sequences. 
Despite the success, the effectiveness of embedding enrichment hinges on the availability of detailed metadata, which may be sparse or entirely missing in real-world scenarios where data privacy matters~\cite{milano2020recommender}. 
Moreover, the increased model complexity due to enriched embeddings might inadvertently result in overfitting, especially in the absence of sufficient supportive data \cite{liu2021noninvasive}.

Different from these works, without requiring detailed metadata and increasing model complexity}, recent studies in SR have aimed to embrace this challenge by integrating effective \emph{data augmentation strategies}~\cite{li2017neural, yuan2020future, wang2021counterfactual} to enrich the context of short sequences. The initial line of research~\cite{xie2022contrastive, zhou2020s3} adopts naive augmentation strategies, \textit{e.g.,} masking, replacement, cropping, and item addition to augment the sequence data. While such approaches are relatively expeditious to implement, they bear the drawback of potentially undermining the inherent sequential properties and underlying correlations within the data \cite{liu2021contrastive,duorec,xie2022contrastive,zhou2023equivariant}. For example, masking or cropping could remove pivotal interactions that serve as context for subsequent choices, leading to an incomplete or distorted view of the user's behavior. Furthermore, these augmentation techniques may exacerbate the challenges posed by the cold-start problem, as short sequences become even more vulnerable to subtle item alterations~\cite{liu2021contrastive}. 

Building upon the challenges observed with naive augmentation, another line of augmentation-based works~\cite{liu2021augmenting} has turned to the \emph{generative paradigm}, aiming to maintain the intrinsic sequential properties while revamp the content of sequences \cite{kenton2019bert,radford2018improving,liu2021augmenting}. 
In particular, within the specific framework of sequential recommendation predicting the next item based on historical user behavior, ASReP~\cite{liu2021augmenting}
harnesses the reverse sequential correlation (\textit{i.e.}, future-to-past) from data to generate pseudo-prior items at the beginning of the original sequences. 
\revise{To equip models with the ability to generate pseudo-prior items, the model undergoes initial pre-training aimed at predicting masked prior items in reverse sequences.
Given the length and contextual augmented sequences, this class of models subsequently fine-tunes the Transformer in the forward direction (\textit{i.e.}, past-to-future) to predict the next items.} 
As shown in Fig. \ref{fig:distribution}, although ASReP has demonstrated improvements in the learning of short sequences, \textcircled{1} the performance on short sequences still lags behind that on longer sequences and \textcircled{2} the similar or worse performance compared to SASRec \cite{kang2018self} in the case of long sequences with $L \geq 20$.
This generative paradigm retains the integrity of the original sequences and merely augments them with the addition of prior ``missing'' items. \round{However, as the augmentation process follows a backward progression (future-to-past), it potentially leads to temporal correlations within the generated sequence that may not be entirely congruent with the original sequence when considered in the forward direction (past-to-future) \cite{liu2021contrastive,duorec,xie2022contrastive,zhou2023equivariant}.} 

In light of the relevance of ASReP to our research, it is imperative to articulate the existing challenges within ASReP and illustrate the distinct solutions our study proposes. Specifically, ASReP's reliance on pseudo prior-items for \revise{naive sequence augmentation is fraught with several limitations:
\textcircled{1} there is no certainty that the augmented sequences preserve user preference integrity. \textcircled{2} as sequences lengthen, the task of capturing long-term dependencies grows more complex, a situation further compounded by the proliferation of items due to reverse augmentation. \textcircled{3} the intention to enrich contextual information might inadvertently undermine the model's representational power by reducing its Shannon entropy. 

\emph{Distinct from prior work, our study endeavors to concurrently and systematically address three complex challenges above}. 
\round{To address the first challenge, our proposed BARec innovates by integrating bidirectional temporal data correlations into the pseudo prior-item generation process, thereby aligning with user preference patterns more effectively. 
This strategy is supported by rigorous theoretical evidence.
Moreover, it maintains bidirectional item correlations, enhancing the accuracy and fidelity of item representation in both forward and reverse sequences.
To effectively resolve the latter two issues, we introduce a knowledge enhancement fine-tuning strategy that dissociates the benefits of data augmentation from the model's representational capabilities. Our approach specifically endeavors to minimize the Kullback-Leibler (KL) divergence between the distributions of augmented and original sequence data. 
This strategy ensures that the representation space uniformly reflects both original and augmented sequences, maintaining the Shannon entropy across varying sequence lengths.}
\emph{Thus, the BARec model we propose represents a significant departure from the existing ASReP framework, distinguished by its unique motivation, structural design, and theoretical underpinnings}.}
\round{We substantiate the effectiveness of BARec through extensive experimentation on four prevalent benchmarks and one real-world large scale benchmarks, with results evidencing marked improvements over a suite of evaluative metrics, thereby confirming the superiority of BARec.}

In summary, the main contributions of this paper are as follows:
\begin{itemize}
    \item We introduce a novel bidirectional temporal data augmentation with pre-training approach that synthesizes pseudo-prior items of high quality while preserving user preferences and enhancing item representations.
    \item We develop a knowledge enhancement fine-tuning technique that decouples the surplus information gained from data augmentation, ensuring consistent mapping of original and augmented sequences into a homogeneous representation space.
    \item We offer a rigorous theoretical analysis of user preference preservation within our framework, affirming that the generated pseudo-prior items maintain user preference congruity with the original forward sequences. 
    \item We conduct comprehensive experimental validation to demonstrate the superior performance of our method over established baselines across various sequence lengths in five benchmark datasets, with notable improvements for extremely short and long sequences. 
\end{itemize}

\section{Related Work}
\subsection{Data Augmentation}\label{sec:data_augmentation}
Data augmentation is an invaluable strategy in deep learning, bolstering dataset size or variety and thus markedly enhancing model performance \cite{shorten2019survey}. Within recommendation systems, diverse data augmentation methodologies have emerged \cite{tan2016improved,yuan2020future,liu2021augmenting,xie2022contrastive,liu2021contrastive,duorec,zhou2023equivariant}. For instance, \cite{tan2016improved} employ sequence preprocessing and embedding dropout to mitigate overfitting and reinforce training robustness. \cite{yuan2020future} suggests augmenting datasets with future interactions to refine training processes.
Generative techniques, which include the use of Generative Adversarial Networks (GANs) or Seq2Seq models, have been advocated to create additional user-item interactions \cite{radford2018improving,kenton2019bert,wang2019enhancing,li2019context}. 
Self-supervised or contrastive learning, another burgeoning approach, applies operations such as cropping, reordering, masking, substituting, and inserting to enrich sequence recommendation tasks \cite{zhou2020s3,xie2022contrastive,liu2021contrastive,duorec,qiu2022contrastive}, thus bolstering model resilience.
\revise{Most recent, several works \cite{wang2023diffusion,liu2023diffusion,yang2024generate} have applied Diffusion Models (DMs), originally utilized in image synthesis, to recommendation systems, aiming to learn user target behaviors through a denoising process.
}
\round{Nonetheless, these methods may not effectively tackle the cold-start dilemma as they could disrupt the sequential integrity of data, and learning from truncated sequences remains sensitive to minor item variations \cite{liu2021contrastive}. 
To address the cold-start issue in short sequences, it has conventionally been approached through leveraging side information 
\cite{saveski2014item,cai2023user}, employing contrastive learning \cite{wei2021contrastive,du2022socially}, transferring knowledge \cite{zhao2020catn,zhu2021transfer}, and utilizing meta-learning frameworks \cite{zheng2021cold,lu2020meta}.
However, these approaches generally require detailed metadata and increase model complexity.
Most recently, \cite{liu2021augmenting} introduced ASReP, a reverse data augmentation technique that appends pseudo-prior items to sequence beginnings by predicting past items in reverse.  
Despite its success, the augmented sequences may disrupt the integrity of user preferences.}

\subsection{Sequential Recommendation}
At the heart of sequential recommendation systems lies the imperative to capture user-item interaction patterns and forecast future items from historical user data. 
Initial methodologies predominantly leveraged Markov chains for pair-wise item transitions 
\cite{rendle2010factorizing,he2016fusing}. 
However, the advent of deep learning has seen a paradigm shift, with neural networks largely supplanting Markov models. 
In particular, Recurrent Neural Networks (RNNs) have gained prominence due to their ability to retain sequential information \cite{hidasi2015session,yu2016dynamic,liu2016context}, and variants with specialized architectures and gating mechanisms have been developed for diverse recommendation scenarios \cite{quadrana2017personalizing,ma2019hierarchical,song2019hierarchical}. 
Enhancements to RNNs, including the incorporation of contextual data \cite{smirnova2017contextual} and novel sampling techniques \cite{hidasi2018recurrent}, have also been explored. 
Convolutional Neural Networks (CNNs) have been introduced to model local item transitions \cite{tang2018personalized, yuan2019simple}, although these still struggle with capturing long-term dependencies \cite{jiang2023adamct}.
More recently, Transformer-based architectures have been increasingly adopted, demonstrating superior performance in sequential recommendation tasks \cite{kang2018self,kang2019recommender, sun2019bert4rec,luo2021stan}. For instance, SASRec repurposes the Transformer mechanism from natural language processing for sequential recommendation \cite{kang2018self}, while BERT4Rec introduces a bidirectional Transformer to further enhance model capabilities \cite{sun2019bert4rec}.
Complementary approaches that integrate side information \cite{vasile2016meta,chen2018collective,liu2021noninvasive}, content features \cite{deldjoo2020recommender,kulkarni2020context}, and \revise{graph neural network (GNNs) frameworks \cite{wu2019session,chang2021sequential,zhang2022dynamic,su2023enhancing} have also been employed to refine sequential recommendation systems and address the nuances of various recommendation tasks.}
\revise{
Furthermore, multi-behavior SR \cite{yuan2022multi,yang2022multi,su2023personalized} leverages various types of user-item interactions to improve performance in predicting target behaviors.
Cross-domain SR \cite{ma2019pi,zheng2022ddghm,guo2021gcn,li2021dual} exploits the overlap of information among users and/or items across different domains, aiming to mitigate challenges associated with cold start and data sparsity.}
\round{In addition, various data augmentation techniques have been developed to address the challenges of data sparsity and short sequences in SR. 
In this study, we propose a novel data augmentation method designed to overcome the limitations of previous approaches in SR, as discussed in Section \ref{sec:data_augmentation}.}

\section{Methodology}
\subsection{Problem Statement}
Let $\mathcal U=\{u_1, u_2, \ldots, u_{|\mathcal U|}\}$ represent a set of users and $\mathcal{I}=\{v_1, v_2, \ldots, v_{|\mathcal{I}|}\}$ denote a collection of items. For each user $u \in \mathcal{U}$, their interaction history with items is chronologically ordered as $\mathcal{S}^{(u)}=[v_1^{(u)}, v_2^{(u)}, \ldots, v_{|\mathcal{S}^{(u)}|}^{(u)}]$, where $v_i^{(u)}$ indicates an interaction with an item from $\mathcal{I}$ at the $i$-th time step. The objective of SR is to predict the subsequent item $v_{|\mathcal{S}^{(u)}|+1}^{(u)}$ with which user $u$ will engage at the $(|\mathcal{S}^{(u)}|+1)$-\textit{th} time step. This is accomplished by estimating the conditional probability distribution $v_{|\mathcal{S}^{(u)}|+1}^{(u)} \sim \mathcal{P}_{\bm\phi}(v \mid \mathcal{S}^{(u)})$, which predicts the likelihood of any item $v \in \mathcal{I}$ being the next interaction, based on the user's historical sequence $\mathcal{S}^{(u)}$ and model parameters $\bm\phi$.
\revise{The summary of the used notations in this paper is illustrated in Table \ref{tab:symbol}}.
\begin{table}[t]
\centering
\caption{Summary of the used notations in this paper, organized from top to bottom based on the order of their initial appearance in the text.}
\label{tab:symbol}
  \resizebox{\linewidth}{!}{
    \begin{tabular}{c|c}
        \hline
         Notation	& Definition \\
        \hline
        \hline
        $\mathcal{U}$ &the set of users\\
        $\mathcal{I}$ &the set of items\\
        $\mathcal{S}^{(u)}$ &the interaction sequence in chronological order of user $u\in\mathcal{U}$\\
        $\mathcal{R}ev(\mathcal{S}{1:n}^{(u)})$ &the interaction sequence in reverse chronological order of user $u\in\mathcal{U}$\\
        $v_i^{(u)}$ &the item in $\mathcal{I}$ that $u$ has interacted with at time step $i$\\
        $d_{model}$ &the latent dimension of encoder\\
        $\mathcal{E}{item}\in \mathbb{R}^{|\bm{\mathcal{I}}|\times d{model}}$ &the item embedding table\\
        $n$ &the maximum sequence length\\
        $\mathcal{P}os \in \mathbb{R}^{n \times d_{model}}$ &the position embedding table\\
        $\tau$ &the temperature for scaled dot-product attention layer\\
        $\mathcal{H}^{(u)}{l-1} \in \mathbb{R}^{n\times d{model}}$ &the $l$-\textit{th} Encoder layer's input\\
        $h^{(u)}{l} \in \mathbb{R}^{n\times d{model}}$ &the output of casual multi-head self-attention (CMHSA) sub-layer\\
        $h$ &the number of attention layers\\
        $\left\{\mathcal{W}_i^\mathcal{Q}, \mathcal{W}_i^\mathcal{K}, \mathcal{W}i^\mathcal{V}, \mathcal{W}i^\mathcal{O}\right\} \in \mathbb{R}^{d{model} \times d{model}/ h}$ &four affine parameters that are unique per layer and attention head\\
        $\zeta_{mask}$ &casual attention masking\\
        $\left\{\mathcal{W}^{(1)}, (\mathcal{W}^{(2)})^T \right\} \in \mathbb{R}^{d_{model} \times 4d_{model}}$ &learnable projection in Position-wise Feed-Forward Network (PFFN)\\
        $\{b^{(1)}, b^{(2)}\} \in \mathbb{R}^{d_{model}}$ &learnable bias parameters in PFFN\\
        $T$ &the transpose of a matrix\\
        $\mathcal{S}{prior}^{(u)}$  &the pseudo-prior items sequence in reverse chronological order of user $u\in\mathcal{U}$\\
        $\bm\Phi$ &the pre-trained model parameters\\
        $v_{-(j+1)}^{(u)} \in \mathcal V$ &the pseudo-prior items generated at $-(j+1)$-\textit{th} time step\\
        $\mathcal{\gamma}_{mask}$ &the masking proportion of last consecutive items in $\mathcal{R}ev(\mathcal{S}_{1:n}^{(u)})$\\
        $t_k^{(u)}$ &the index of items in $\mathcal{R}ev(\mathcal{S}_{1:n}^{(u)})$\\
        $\mathcal{T}^{(u)}_{pseudo}$ &the pseudo-prior items of labels\\
        $\mathcal{L}_{pseudo}$  &the objective loss of the pseudo-prior items generation\\
        $\mathcal{\gamma}_{prefer}$ &the user preference proportion of last consecutive items in the original data $\mathcal{S}_{1:n}^{(u)}$\\
        $f_k^{(u)}$ &the index of items in $\mathcal{S}_{1:n}^{(u)}$\\
        $\mathcal{F}^{(u)}_{prefer}$ &the user preference modeling\\
        $\lambda$ &the trade-off between forward predictions and reverse generation\\
        $\mathcal{K}$  &the number of the pseudo-prior items expected to be generated \\
        $\mathcal{M}$ &the length threshold at which a sequence is worth augmenting with a pseudo-prior item.\\
        $\mathcal{A}ug(\mathcal{S}_{1:n}^{(u)})$ &the augmented sequence data\\
        $\mathcal{\eta}_{gold}$  &the proportion of gold label\\
        $clip_k$ &the number of gold label in the shift-one-step of the input sequence\\
        $\bm\varphi$ &the model parameters of fine-tuning\\
        $\mathcal{A}ug(\mathcal{S}{1:n}^{(u)}){\leq t}$ &the contextual input sequence at $t$-\textit{th}\\
        $\mathcal{P}(\cdot\mid\cdot)$ &the conditional probability distribution\\
        $H(\cdot)$ &the Shannon entropy of the sequence data\\
        $\mathcal{L}_{KL}(\bm\varphi)$ &the bidirectional Kullback-Leibler (KL) divergence\\
        $\mathcal{L}_{FT}^{final}$ &the final fine-tuning loss\\
        $\alpha$ &the trade-off between next item prediction and model knowledge enhancement.\\
    \bottomrule
    \end{tabular}
    }
\end{table}

\begin{figure*}[t]
	\centering
	\includegraphics[width=0.7\linewidth]{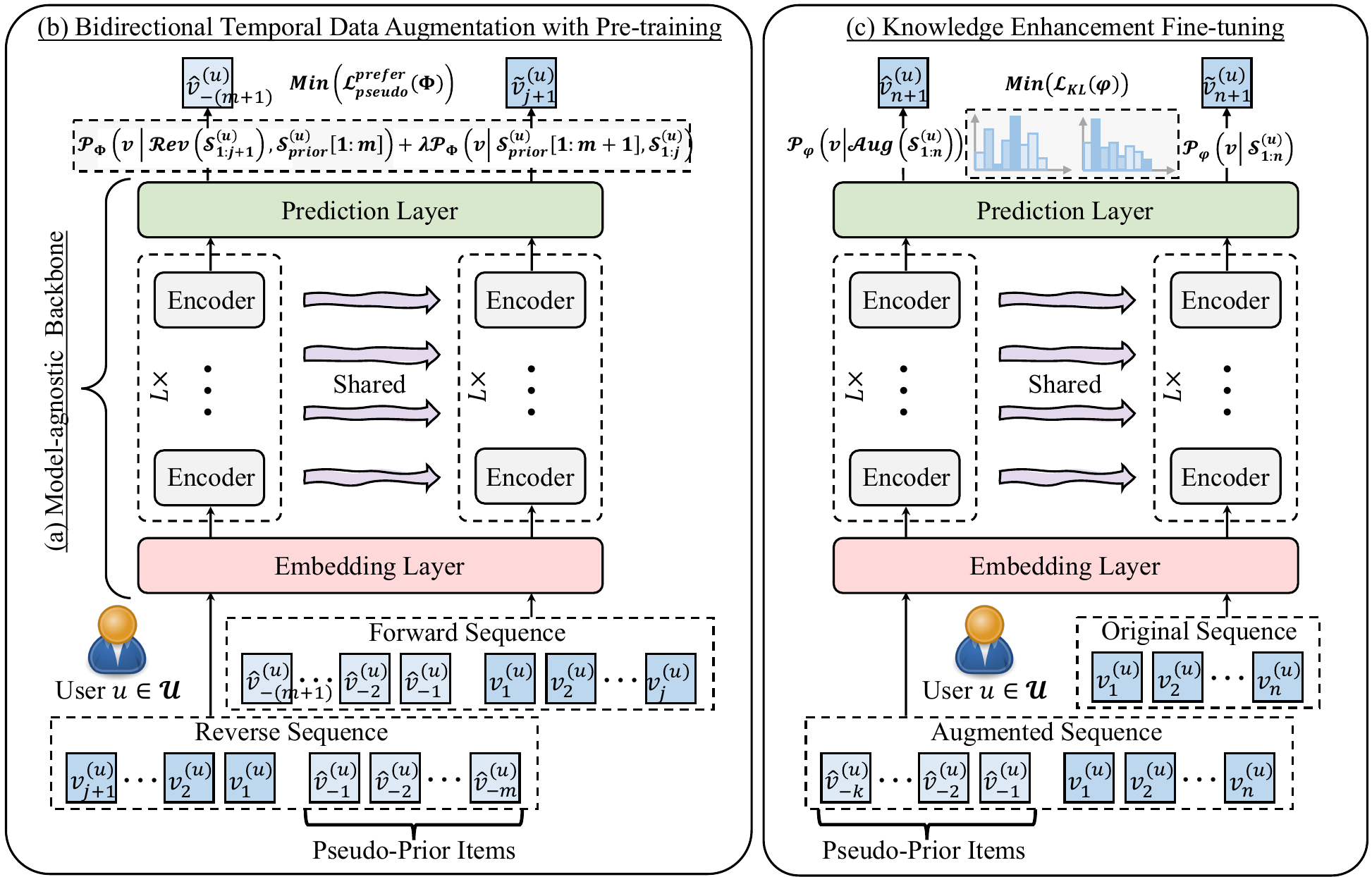}
	\caption{
The architecture of our proposed BARec framework encompasses three core components: (a) a model-agnostic backbone featuring an embedding layer, $L\times$ Encoder layers, and a prediction layer; (b) a bidirectional temporal data augmentation strategy, coupled with pre-training, designed to generate high-fidelity pseudo-prior items that retain user preferences and enhance the semantic interrelations among items; (c) a knowledge-enhanced fine-tuning phase tailored for downstream tasks, specifically, sequential recommendation.
 }
\label{fig:arch}
\vspace{-5mm}
\end{figure*}

\subsection{Framework Pipeline}
In this work, we introduce a versatile data augmentation strategy that is \emph{model-agnostic} and employs a combined pre-training and fine-tuning paradigm. The comprehensive framework is depicted in Figure \ref{fig:arch}.
The subsequent sections provide an in-depth illustration of each integral component of BARec.

\subsection{Model-agnostic Backbone}
\subsubsection{Embedding Layer}
The embedding layer in a standard Transformer architecture \cite{vaswani2017attention} comprises two components: item and positional embeddings. Specifically, the item embedding space $\mathcal{E}{item}\in \mathbb{R}^{|\bm{\mathcal{I}}|\times d_{model}}$ projects discrete item identifiers $v_i \in \mathbb{R}$ into a continuous $d_{model}$-dimensional latent space, resulting in dense vector representations $e_i \in \mathbb{R}^{d_{model}}$.
Consistent with the methodology presented in \cite{liu2021augmenting}, we maintain a positional embedding matrix $\mathcal{P}os \in \mathbb{R}^{n \times d_{model}}$, where $n$ denotes the maximum permissible sequence length and $d_{model}$ aligns with the dimensionality of the item embeddings. For sequence processing, sequences $\mathcal{S}^{(u)}$ exceeding length $n$ are truncated, while shorter sequences are padded with zeros to achieve the fixed length $n$ as delineated in \cite{kang2018self}.
The embedding layer therefore maps an input sequence $\mathcal{S}_{1:n}^{(u)}=[v_1^{(u)}, v_2^{(u)}, ..., v_n^{(u)}] \in \mathbb{R}^{n}$ to a joint embedding space $\mathcal{E}_{(item,pos)}(\mathcal{S}^{(u)}_{1:n})= [\mathcal{E}_{item}(\mathcal{S}^{(u)}_{1:n}) + \mathcal{P}os(\mathcal{S}^{(u)}_{1:n})]\in \mathbb{R}^{n \times d_{model}}$, effectively capturing both the item-specific features and their positional context within the sequence.

\subsubsection{Encoder Layer}
\textit{Sub-layer 1: Casual Multi-Head Self-Attention.} 
The causal multi-head self-attention (CMHSA) component of our model is a composite of a multi-head self-attention mechanism with a causal masking strategy, followed by layer normalization of the latent sequential representations. 
This sub-layer is structured to harness diverse contextual patterns across $h$ attention heads, each utilizing scaled dot-product attention with a scaling factor $\tau = \sqrt{d_{model}/h}$, as suggested in prior works \cite{vaswani2017attention,sun2019bert4rec,kang2018self,liu2021augmenting}. This scaling mitigates the risk of disproportionately large inner products and correspondingly diminished gradients, thus fostering a more balanced attention landscape. Formally, 
\begin{equation}
\begin{aligned}
h_{l}^{(u)} = \operatorname{CMHSA}(\mathcal{H}_{l-1}^{(u)})=\operatorname{LN}(\operatorname{DP}(\operatorname{Concat}\{\text{Head}_i\}_{i=1}^{h}\mathcal{W}^\mathcal{O})),
\end{aligned} 
\end{equation}
\begin{equation}
\begin{aligned}
\text{Head}_i = \text{Att}(\underbrace{\mathcal{H}_{(u)}^{l-1}\mathcal{W}_i^\mathcal{Q}}_{\mathcal{Q}},\underbrace{\mathcal{H}_{(u)}^{l-1}\mathcal{W}_i^\mathcal{K}}_{\mathcal{K}}, \underbrace{\mathcal{H}_{(u)}^{l-1}\mathcal{W}_i^\mathcal{V}}_\mathcal{V}), 
\end{aligned}     
\end{equation}
\begin{equation}
\begin{aligned}
\text{Att}(\mathcal{Q},\mathcal{K},\mathcal{V})=\mathrm{softmax}(\mathcal{Q}\mathcal{K}^\mathcal{T}/{\tau} + \zeta_{mask})\mathcal{V},
\end{aligned}   
\end{equation}
\begin{equation}
\begin{aligned}
    \zeta_{mask} = \Big(c_{ij}\Big)_{n\times n} = \Big(\mathbb{I}(i\ge j)\Big)_{n\times n} = 
    \begin{cases}
    0 & \text{for $i \ge j$ } \\
    -\infty & \text{otherwise}
    \end{cases},
\end{aligned}
\end{equation}
where $\mathcal{H}^{(u)}_{l-1} \in \mathbb{R}^{n\times d_{model}}$ denotes the input to the $l$-\textit{th} encoder layer, and $h^{(u)}_{l} \in \mathbb{R}^{n\times d_{model}}$ represents the CMHSA sub-layer's output with $\mathcal{H}^{(u)}_{0}=\mathcal{E}(\mathcal{S}^{(u)}_{1:n})$ being the initial embedding sequences. 
LN and DP correspond to layer normalization and dropout, respectively. 
The number of distinct attention heads is denoted by $h$. The sets $\left\{\mathcal{W}_i^\mathcal{Q}, \mathcal{W}_i^\mathcal{K}, \mathcal{W}_i^\mathcal{V}, \mathcal{W}_i^\mathcal{O}\right\} \in \mathbb{R}^{d_{model} \times d_{model}/ h}$ contain the affine transformation parameters for each attention head, transforming the Query $\mathcal{Q}$, Key $\mathcal{K}$, Value $\mathcal{V}$, and the attention sub-layer's output. 
Causal attention masking $\zeta_{mask}$ is implemented by setting the lower triangular part to 0 and the remaining elements to $-\infty$, ensuring that each item attends only to its predecessors and itself, thereby maintaining alignment with the next-item prediction task.

\textit{Sub-layer 2: Position-wise Feed-Forward Network.} 
A Position-wise Feed-Forward Network (PFFN) is utilized subsequent to the CMHSA sub-layer to refine the sequence embeddings at each position $i$ and encode more intricate feature representations. 
The PFFN consists of a pair of linear transformations interspersed with a ReLU activation function, in line with \cite{vaswani2017attention,kang2018self,liu2021augmenting}.
\begin{equation}
\begin{aligned}
\text{PFFN}(h^{(u)}_{l})=\left(\text{Concat}\left\{\text{FFN}(h^{(u)}_{l}[i])^T\right\}_{i=1}^{n}\right)^T,
\end{aligned}   
\end{equation}
\begin{equation}
\begin{aligned}
\text{FFN}(h^{(u)}_{l}[i])=\text{ReLU}(h^{(u)}_{l}[i]\mathcal{W}^{(1)}+b^{(1)})\mathcal{W}^{(2)}+b^{(2)},
\end{aligned} 
\end{equation}
where the projection matrices $\left\{\mathcal{W}^{(1)}, (\mathcal{W}^{(2)})^T \right\} \in \mathbb{R}^{d_{model} \times 4d_{model}}$ and bias vectors $\{b^{(1)}, b^{(2)}\} \in \mathbb{R}^{d_{model}}$ are parameters learned during training and are consistent across all positions. Here, $T$ denotes the transpose operation on a matrix.

\subsection{Bidirectional Temporal Data Augmentation with Pre-training}
\label{sec:pre-training}
Initially, we invert the user-specific sequence $\mathcal{S}_{1:n}^{(u)}=[v_1^{(u)}, v_2^{(u)}, ..., v_n^{(u)}] \in \mathbb{R}^{n}$ to obtain $\mathcal{R}ev(\mathcal{S}_{1:n}^{(u)})=[v_n^{(u)}, v_{n-1}^{(u)}, ..., v_1^{(u)}] \in \mathbb{R}^{n}$. 
This reversed sequence is input into the Encoder to facilitate the autoregressive prediction of subsequent items, a process we term pseudo-prior item generation, which yields $\mathcal{S}_{prior}^{(u)}=\{v_{-1}^{(u)}, v_{-2}^{(u)}, ..., v_{-|\mathcal{S}_{prior}^{(u)}|}^{(u)}\}$, with each $v_{-j}^{(u)}$ belonging to the item set $\mathcal{V}$. 
Formally, we express this as: 
\begin{equation}
\begin{aligned}
v_{-(j+1)}^{(u)} \sim \mathcal{P}_{\bm\Phi}\Big(v \mid \mathcal{R}ev(\mathcal{S}_{1:n}^{(u)}), \mathcal{S}_{prior}^{(u)}[1:j]\Big),
\end{aligned}  
\end{equation}
where $\bm\Phi$ represents the pre-trained model parameters, $\mathcal{R}ev(\mathcal{S}_{1:n}^{(u)})$ is the reversed sequence, $v_{-(j+1)}^{(u)} \in \mathcal V$ denotes the pseudo-prior items generated at the $-(j+1)$-th time step, and $\mathcal{S}_{prior}^{(u)}[1:j]$ includes the pseudo-prior items generated to date.

During pre-training, we construct pseudo-prior item labels by masking a fraction $\mathcal{\gamma}_{mask}$ of the final consecutive items $\mathcal{T}^{(u)}_{pseudo}=\{t^{(u)}_1, t^{(u)}_2, ..., t^{(u)}_{\lfloor n\mathcal{\gamma}_{mask}\rfloor}\}$, where $t_k^{(u)}$ indexes the last consecutive items in $\mathcal{R}ev(\mathcal{S}_{1:n}^{(u)})$. 
The model is then pre-trained to predict these masked items, with the objective loss for pseudo-prior item generation denoted as:  
\begin{equation}
\begin{aligned}
\mathcal{L}_{pseudo} = -\sum_{u\in\mathcal{U}}\sum_{t\in\mathcal{T}^{(u)}_{pseudo}}\log \mathcal{P}_{\bm\Phi}\Big(\mathcal{S}_{-t}^{(u)}\mid \\ 
\mathcal{R}ev(\mathcal{S}_{n-\lfloor n\mathcal{\gamma}_{mask}\rfloor:n}^{(u)}), \mathcal{S}_{prior}^{(u)}[1:t-1]\Big).
\end{aligned}   
\end{equation}
This reverse learning strategy enables the Encoder to discern item correlations in the reverse direction \cite{sun2019bert4rec,liu2021augmenting}, capturing a user's temporal preference pattern in reverse.
\round{However, as discussed in Sec. \ref{sec:analysis}, straightforward pre-training on $\mathcal{R}ev(\mathcal{S}_{1:n}^{(u)})$ may yield diminished performance in downstream tasks, such as next-item prediction due to a potential mismatch in user preference patterns. 
To address this, we introduce a novel bidirectional temporal pre-training. The fundamental premise is that pseudo-prior items generated in reverse should align with forward user preferences, thus enhancing pre-trained model effectiveness for downstream tasks.}

Specifically, the pseudo-prior item $v_{-(j+1)}^{(u)} \in \mathcal V$ produced at the $-(j+1)$-th time step should maintain consistency with the user's original preference pattern in $\mathcal{S}_{1:n}^{(u)}$. To model user preference, we utilize a proportion $\mathcal{\gamma}_{prefer}$ of the final consecutive items $\mathcal{F}^{(u)}_{prefer}=\{f^{(u)}_1, f^{(u)}_2, ..., f^{(u)}_{\lfloor n\mathcal{\gamma}_{prefer}\rfloor}\}$, with $f_k^{(u)}$ indexing the last consecutive items in $\mathcal{S}_{1:n}^{(u)}$. 
The pre-trained model subsequently uses the pseudo-prior item $v_{-(j+1)}^{(u)}$ to make forward predictions of items $\mathcal{F}^{(u)}_{prefer}$, reflecting the user preference pattern. The pre-training objective loss $\mathcal{L}_{pseudo}$ is enhanced by:  
\begin{equation}
\begin{aligned}
\label{eq:pre-training}
\mathcal{L}_{pseudo}^{prefer} &= -\sum_{u\in\mathcal{U}}\sum_{t\in\mathcal{T}^{(u)}_{pseudo}}\\\Big[ 
\log \mathcal{P}_{\bm\Phi}&\Big(\mathcal{S}_{-t}^{(u)}\mid \mathcal{R}ev(\mathcal{S}_{n-\lfloor n\mathcal{\gamma}_{mask}\rfloor:n}^{(u)}), \mathcal{S}_{prior}^{(u)}[1:t-1]\Big) \\ 
+ \lambda\sum_{f\in\mathcal{F}^{(u)}_{prefer}}&\log \mathcal{P}_{\bm\Phi}\Big(\mathcal{S}_{f}^{(u)}\mid 
\mathcal{S}_{prior}^{(u)}[1:f], \mathcal S_{1:n-\lfloor n\mathcal{\gamma}_{prefer}\rfloor}^{(u)}\Big)\Big],
\end{aligned}  
\end{equation}
where $\lambda$ is a hyper-parameter that balances the trade-off between forward predictions and reverse generation. 
Through this bidirectional temporal data augmentation, the Encoder ensures that pseudo-prior items generated in reverse are consistent with forward user preferences, thereby improving the model's effectiveness for downstream tasks.

\textbf{Autoregressive Item Generation}. 
Upon completion of pre-training, the model is primed for pseudo-prior item generation. Our analysis of benchmark datasets reveals a prevalence of short sequences, with long sequences being relatively infrequent (refer to Fig. \ref{fig:distribution}). In light of this, we introduce two hyper-parameters, $\mathcal{K}$ and $\mathcal{M}$, to denote the number of pseudo-prior items to be generated and the sequence length threshold for augmentation, respectively. Through autoregressive generation, as delineated in previous studies \cite{rendle2010factorizing,yuan2019simple,liu2021augmenting} (detailed in Algorithm \ref{al:generate}), we acquire enriched sequence data as: 
\begin{equation}
\begin{aligned}
    \mathcal{A}ug(\mathcal{S}_{1:n}^{(u)}) &=  
    \begin{cases}
    [\mathcal{S}_{prior}^{(u)}, \mathcal{S}_{1:n}^{(u)}] & \text{for $|\mathcal{S}_{1:n}^{(u)}| < \mathcal{M}$} \\
    \mathcal{S}_{1:n}^{(u)} & \text{otherwise}
    \end{cases},\\
    \mathcal{S}_{prior}^{(u)}&=\{v_{-1}^{(u)}, v_{-2}^{(u)}, ..., v_{-min(\mathcal{K}, \mathcal{M})}^{(u)}\}.
\end{aligned}
\end{equation}

\subsection{Knowledge Enhancement Fine-tuning}
\label{sec:fine-tuning}
In this section, we fine-tune the above pre-trained model, $\mathcal{P}_{\bm\varphi}(\cdot)$, utilizing augmented sequential data $\mathcal{A}ug(\mathcal{S}_{1:n}^{(u)})$ to address the SR task. 
Our approach employs a modified version of the common \emph{shift-one-step} labeling technique \cite{kang2018self,sun2019bert4rec}, selecting the most recent subset of gold labels, $clip_k = \lfloor|\mathcal{A}ug(\mathcal{S}_{1:n}^{(u)})|\mathcal{\eta}_{gold}\rfloor$, where $\mathcal{\eta}_{gold}$ represents a predefined proportion. This modification addresses the limitations posed by short sequences and mitigates the \emph{cold-start issue}, as discussed in \cite{liu2021augmenting}. The fine-tuning loss function is defined as follows: 
\begin{equation}
\begin{aligned}
\label{eq:ft}
    \mathcal{L}_{FT}(\bm\varphi) = -\sum_{u=1}^{|\mathcal{U}|}\sum_{t=n-clip_k+1}^{|\mathcal{A}ug(\mathcal{S}_{1:n}^{(u)})|}\Big[\log\mathcal{P}_{\bm\varphi}\Big(v_{t+1}^{(u)}\mid \mathcal{A}ug(\mathcal{S}_{1:n}^{(u)})_{\leq t}\Big)\\ +
    \sum_{v_j^{(u)}\notin \mathcal{A}ug(\mathcal{S}_{1:n}^{(u)})}\log \Big(1-\mathcal{P}_{\bm\varphi}\Big(v_{j}^{(u)}\mid \mathcal{A}ug(\mathcal{S}_{1:n}^{(u)})_{\leq t}\Big)\Big)\Big],
\end{aligned}
\end{equation}
where $\bm\varphi$ represents the parameters updated during fine-tuning and $\mathcal{A}ug(\mathcal{S}_{1:n}^{(u)})_{\leq t}$ signifies the contextual input at the $t$-th time step. 

\round{Although our data augmentation preserves the intrinsic item correlations, it enhances the contextual richness of the source data. Entropy analysis below reveals that the Shannon entropy of augmented data is expectedly lower than that of the original sequences, which may cause model representational sluggishness and potential overfitting \cite{gray2011entropy}: 
\begin{equation}
\begin{aligned}
\label{eq:information_theory}
    H(\mathcal{P}_{\bm\varphi})_{aug} &< H(\mathcal{P}_{\bm\varphi})_{org},\\
    H(\mathcal{P}_{\bm\varphi})_{aug} &= \mathbb{E}_{v_{t+1}\sim \mathcal{P}_{\bm\varphi}}\Big[-\log\mathcal{P}_{\bm\varphi}\Big(v\mid \mathcal{A}ug(\mathcal{S}^{(u)})_{\leq t}\Big)\Big],\\
    H(\mathcal{P}_{\bm\varphi})_{org} &= \mathbb{E}_{v_{t+1}\sim \mathcal{P}_{\bm\varphi}}\Big[-\log\mathcal{P}_{\bm\varphi}\Big(v\mid \mathcal{S}^{(u)}_{\leq t}\Big)\Big],\\
    \mathcal{P}_{\bm\varphi}\Big(v\mid \mathcal{A}&ug(\mathcal{S}^{(u)})_{\leq t}\Big) \ge \mathcal{P}_{\bm\varphi}\Big(v\mid \mathcal{S}^{(u)}_{\leq t}\Big).
\end{aligned}
\end{equation}}

\round{To mitigate this risk, we decouple the augmented information from the model's representational learning by aligning the probability distributions of the augmented $\mathcal{P}_{\bm\varphi}\Big(v\mid \mathcal{A}ug(\mathcal{S}^{(u)})_{\leq t}\Big)$ and original $\mathcal{P}_{\bm\varphi}\Big(v\mid \mathcal{S}^{(u)}_{\leq t}\Big)$ sequences. 
We refer to this process as \emph{model knowledge enhancement}.
A multitude of metrics exists to achieve this objective; however, 
KL-divergence offers a nuanced and efficient way to measure the divergence between the true distribution of user actions and the distribution predicted by a SR model. 
Its sensitivity to the probabilistic nature of distributions, combined with its advantages in handling the sequential nature of user behavior, make it an excellent choice for optimizing SR.
Therefore, we employ the bidirectional Kullback-Leibler (KL) divergence in this study.
Formally,
}
\begin{equation}
\begin{aligned}
\label{eqn:kl}
    \mathcal{L}_{KL}(\bm\varphi) &= \frac{1}{2}\Big(
        \mathcal{D}_{KL}\Big(\mathcal{P}_{\bm\varphi}\Big(v\mid \mathcal{A}ug(\mathcal{S}^{(u)})_{\leq t}\Big) \mid\mid \mathcal{P}_{\bm\varphi}\Big(v\mid \mathcal{S}^{(u)}_{\leq t}\Big)\Big) \\
        & + \mathcal{D}_{KL}\Big(\mathcal{P}_{\bm\varphi}\Big(v\mid \mathcal{S}^{(u)}_{\leq t}\Big) \mid\mid \mathcal{P}_{\bm\varphi}\Big(v\mid \mathcal{A}ug(\mathcal{S}^{(u)})_{\leq t}\Big)\Big)\Big),\\
         = - \frac{1}{2}\Big(\sum &\mathcal{P}_{\bm\varphi}\Big(v\mid \mathcal{A}ug(\mathcal{S}^{(u)})_{\leq t}\Big) \log \Big( \frac{\mathcal{P}_{\bm\varphi}\Big(v\mid \mathcal{S}^{(u)}_{\leq t}\Big)}{\mathcal{P}_{\bm\varphi}\Big(v\mid \mathcal{A}ug(\mathcal{S}^{(u)})_{\leq t}\Big)}\Big)\\
        + \sum &\mathcal{P}_{\bm\varphi}\Big(v\mid \mathcal{S}^{(u)}_{\leq t}\Big) \log \frac{\mathcal{P}_{\bm\varphi}\Big(v\mid \mathcal{A}ug(\mathcal{S}^{(u)})_{\leq t}\Big)}{\mathcal{P}_{\bm\varphi}\Big(v\mid \mathcal{S}^{(u)}_{\leq t}\Big)}\Big).
\end{aligned}
\end{equation}  
The composite fine-tuning loss integrates this enhancement strategy and is given by:
\begin{equation}
\begin{aligned}
\label{eq:fine-tuning}
    \mathcal{L}_{FT}^{final} = \mathcal{L}_{FT} + \alpha \cdot \mathcal{L}_{KL},
\end{aligned}    
\end{equation}
where $\alpha$ is a hyper-parameter that calibrates the balance between next-item prediction accuracy and knowledge enhancement.


\begin{algorithm}[t]
\caption{The BARec Algorithm for Autoregressive Generation of Pseudo-prior Items}
\label{al:generate}
\SetAlgoLined
\SetKwInOut{Input}{\textbf{Input}}
\Input{The original sequences $\mathcal{S}_{1:n}^{(u)}$.\\
The reversed original sequences $\mathcal{R}ev(\mathcal{S}_{1:n}^{(u)})$.\\
The pre-trained model $\mathcal{P}_{\bm\Phi}$ in Section 3.4.\\
}
\SetKwInOut{Output}{\textbf{Output}} 
\Output{The augmented sequences $\mathcal{A}ug(\mathcal{S}_{1:n}^{(u)})$.}
\BlankLine
Initialize the number of the pseudo-prior items $\mathcal{K}$.\\
Initialize the length threshold $\mathcal{M}$.\\
Initialize model parameters with pre-training $\bm\Phi$.\\
Initialize pseudo-prior items $\mathcal{S}_{prior}^{(u)}=\emptyset$.\\
Initialize augmented sequences $\mathcal{A}ug(\mathcal{S}_{1:n}^{(u)})\gets[\mathcal{S}_{1:n}^{(u)}]$.
\BlankLine
\For{$k$ in $[1, \mathcal{K}]$}{
    \eIf{Sequence Length $|\mathcal{A}ug(\mathcal{S}_{1:n}^{(u)})| < \mathcal{M}$}{
    \underline{\textbf{Greedy Generation}} \\
    Model predicts the next item $v_{-k}^{(u)} = \mathop{\operatorname*{arg\,max}}\limits_v 
    \mathcal{P}_{\bm\Phi}\Big(v \mid \mathcal{R}ev(\mathcal{S}_{1:n}^{(u)}), \mathcal{S}_{prior}^{(u)}\Big)$.\\
    Append the $v_{-k}^{(u)}$ to the pseudo-prior items $\mathcal{S}_{prior}^{(u)}=\{v_{-1}^{(u)}, v_{-2}^{(u)}, ..., v_{-(k-1)}^{(u)}\} \cup v_{-k}^{(u)}$.\\
    \underline{\textbf{Concatenate Sequences}} \\
    Update the augmented sequences $\mathcal{A}ug(\mathcal{S}_{1:n}^{(u)})=[\mathcal{S}_{prior}^{(u)},\mathcal{S}_{1:n}^{(u)}$].
    }
    {Break the for loop}
    }
\Return Augmented sequences $\mathcal{A}ug(\mathcal{S}_{1:n}^{(u)})$
\end{algorithm}
\section{Theoretical Analysis} 
\label{sec:analysis}
\round{In this section, we present a rigorous derivation utilizing Bayes' theorem to establish that forward sequential correlation does not inherently imply consistency with reverse sequential correlation.} 
Let us consider $\mathcal{A}$ and $\mathcal{B}$ as temporally ordered item sequences with which a user interacts. The forward correlation, defined as the conditional probability $\mathcal{P}(\mathcal{B}\mid\mathcal{A})$, quantifies the likelihood of interaction with sequence $\mathcal{B}$ subsequent to $\mathcal{A}$. Conversely, the reverse correlation is characterized by the conditional probability $\mathcal{P}(\mathcal{A}\mid\mathcal{B})$, reflecting the probability of engaging with sequence $\mathcal{A}$ after $\mathcal{B}$: 

\begin{equation}
\begin{aligned}
\mathcal{P}(\mathcal{A}\mid\mathcal{B})\coloneqq \frac{\mathcal{P}(\mathcal{A})}{\mathcal{P}(\mathcal{B})} \cdot \mathcal{P}(\mathcal{B}\mid\mathcal{A}), 
\end{aligned}   
\end{equation}
where $\mathcal{P}(\mathcal{B}) \neq 0$. Thus, we have 
\begin{equation}
\begin{aligned}
    \mathcal{P}(\mathcal{A}\mid\mathcal{B}) &\varpropto \mathcal{P}(\mathcal{A}) \cdot \mathcal{P}(\mathcal{B}\mid\mathcal{A}),\\
    \mathcal{P}(\mathcal{A}\mid\mathcal{B}) = \mathcal{P}(\mathcal{B}\mid\mathcal{A}) &\Rightarrow \mathcal{P}(\mathcal{A})=\mathcal{P}(\mathcal{B}),\\
    \mathcal{P}(\mathcal{A})=\mathcal{P}(\mathcal{B}) &\Rightarrow \mathcal{P}(\mathcal{A}\mid\mathcal{B}) = \mathcal{P}(\mathcal{B}\mid\mathcal{A}),\\
    \mathcal{P}(\mathcal{A}\mid\mathcal{B}) = \mathcal{P}(\mathcal{B}\mid\mathcal{A}) &\Leftrightarrow \mathcal{P}(\mathcal{A})=\mathcal{P}(\mathcal{B}),
\end{aligned}
\end{equation} 
which suggests an assumption of concurrent occurrence of item sets $\mathcal{A}$ and $\mathcal{B}$, an assumption that often does not hold in practical contexts. 
Consequently, pre-training on reversed data without consideration of this discrepancy may yield diminished performance in downstream tasks, such as next-item prediction, stemming from a misalignment with genuine user preference patterns.

\round{In response to the identified challenge, we introduce a novel bidirectional temporal pre-training strategy. To validate its effectiveness, we initially employ $\mathcal{A}$ and $\mathcal{B}$ to generate 
pseudo-prior sequences $\mathcal{C} \sim \mathcal{P}(\mathcal{C}\mid\mathcal{B},\mathcal{A})$ in reverse, and to predict subsequent sequences $\mathcal{D} \sim \mathcal{P}(\mathcal{D}\mid\mathcal{A},\mathcal{B})$ in forward. 
Our objective is to utilize the triplet $(\mathcal{C}, \mathcal{A}, \mathcal{B})$ to enhance the predictive likelihood of $\mathcal{D}$, such that $\mathcal{P}(\mathcal{D}\mid\mathcal{C},\mathcal{A},\mathcal{B}) \ge \mathcal{P}(\mathcal{D}\mid\mathcal{A},\mathcal{B})$.
Throughout the bidirectional temporal pre-training process, we meticulously ensure that the generation of pseudo-prior sequences does not compromise or alter the intrinsic user preferences.}
Formally, 
\begin{equation}
\begin{aligned}
\mathcal{Y}^* \coloneqq &\mathop{\operatorname*{arg\,max}}\limits_\mathcal{Y} \left(\mathcal{P}_{\bm\Phi}(\mathcal{Y}\mid\mathcal{X},\mathcal{B},\mathcal{A}) + \mathcal{P}_{\bm\Phi}(\mathcal{X}\mid\mathcal{Y},\mathcal{A}, \mathcal{B})\right) \\
&\mathrm{s.t. } \quad \mathcal{P}_{\bm\Phi}(\mathcal{X}\mid\mathcal{A},\mathcal{B}) \leq \mathcal{P}(\mathcal{X}\mid\mathcal{Y},\mathcal{A},\mathcal{B}),
\end{aligned} 
\end{equation}
where $\bm\Phi$ denotes the parameters of the pre-trained model, $\mathcal{Y}^*$ represents the optimally generated pseudo-prior sequences that temporally precede $\mathcal{A}$, and $\mathcal{X}$ signifies the chronologically ordered sequence of user interactions following $\mathcal{B}$.

In particular, setting $\mathcal{Y}\coloneqq\mathcal{C}$ and $\mathcal{X}\coloneqq\mathcal{D}$, the aforementioned inequality (our objective) is proven.  
\begin{equation}
\begin{aligned}
\mathcal{Y}\coloneqq\mathcal{C}, \mathcal{X}\coloneqq\mathcal{D} \Rightarrow
\mathcal{P}(\mathcal{D}\mid\mathcal{C},\mathcal{A},\mathcal{B}) \ge \mathcal{P}(\mathcal{D}\mid\mathcal{A},\mathcal{B})
\end{aligned} 
\end{equation}

\section{Experiments}
In this section, we devote to empirically addressing four pivotal Research Questions (RQs) 
\begin{itemize}
    \item \textbf{RQ1:} How does BARec perform compared to state-of-the-art baselines?
    \item \textbf{RQ2:} How does BARec perform on sequences of varying lengths, especially for very short and very long sequences?
    \item \textbf{RQ3:} How do various data augmentation strategies impact the performance?
    \item \textbf{RQ4:} How do the individual components of BARec contribute to its performance?
    \item \textbf{RQ5:} How sensitive is BARec to different hyperparameter settings?
    \item \textbf{RQ6:} How does BARec perform in real-world large-scale recommendation scenarios?
\end{itemize}  
\subsection{Experimental Settings}
\subsubsection{Benchmark Datasets} 
To ensure a fair comparison with existing literature and to validate the efficacy of our proposed model, 
we employ the Amazon datasets\footnote{\scriptsize{\url{https://jmcauley.ucsd.edu/data/amazon}}} introduced by \cite{kang2018self}, which comprise extensive collections of product reviews from \emph{Amazon.com}, organized by the top product categories. For our study, we focus on the Amazon Beauty (5-core), Cell Phones and Accessories (5-core), \round{and Sports and Outdoors (5-core) three categories}, widely acknowledged as standard benchmarks for sequence recommendation research, as corroborated by numerous studies \cite{chen2022intent,liu2021augmenting,fan2022sequential,xie2022contrastive}. 
\round{In addition, we adopt Yelp, a renowned platform for business recommendations encompassing restaurants, bars, beauty salons, among others. 
Following \cite{zhou2020s3,xie2022contrastive}, we utilize transaction records after January 1st, 2019. 
To accurately mirror real-world recommendation scenarios, we employ the large-scale and publicly accessible Tenrec \cite{yuan2022tenrec} dataset, a novel dataset designed for recommender systems that captures diverse user feedback across four distinct recommendation scenarios.
In this study, we specifically focus on the QK-video category.}
Adhering to established preprocessing methodologies \cite{kang2018self,sun2019bert4rec,liu2021augmenting}, we process ratings and reviews as indicators of user-item interactions, organize these interactions by user ID, and chronologically order them to construct interaction sequences for individual users.
The comprehensive dataset statistics are provided in Table \ref{tab:stat}.

\subsubsection{Baselines Models \& Implementation Details}
For comparative analysis, we selected a suite of contemporary superlative sequence recommendation (SR) baselines, comprising 
SASRec \cite{kang2018self}, BERT4Rec \cite{sun2019bert4rec}, \round{SRGNN \cite{wu2019session}, CL4SRec \cite{xie2022contrastive}}, ASReP \cite{liu2021augmenting}. 
We also report results from several model variants we investigated. Notably, variants with subscript ``$\text{w/o PT}$'' indicate models trained de novo on augmented sequences, eschewing the fine-tuning approach.
To ensure equitable comparison, we engaged in meticulous hyper-parameter optimization to ascertain the models optimal performance. Specifically, the optimal hyper-parameters for our proposed BARec model
are delineated as follows:
\textbf{Beauty Category}: Maximum sequence length $n=100$, encoder layers $L=2$, model dimension $d_{model}=128$, attention heads $h=4$, dropout rate $d=0.7$, $\ell_2$ regularization $L2=0.0$, sequence length threshold $\mathcal{M}=18$, pseudo-prior items $\mathcal{K}=15$, masked pseudo-items $|\mathcal{T}^{(u)}{pseudo}|=1$, preferred user items $|\mathcal{F}^{(u)}{prefer}|=1$, trade-off coefficient $\lambda=0.4$, clipped gold labels $clip_k=12$, and balance coefficient $\alpha=1.0$.
\textbf{Phones Category}: $n=100, L=2, d_{model}=32, h=2, d=0.5, L2=0.0, \mathcal{M}=18, \mathcal{K}=17, |\mathcal{T}^{(u)}{pseudo}|=1, |\mathcal{F}^{(u)}{prefer}|=1, \lambda=0.3, clip_k=12, \alpha=0.2$.
\round{\textbf{Sports and Outdoors}: $n=50, L=2, d_{model}=32, h=2, d=0.5, L2=0.0, \mathcal{M}=16, \mathcal{K}=17, |\mathcal{T}^{(u)}{pseudo}|=1, |\mathcal{F}^{(u)}{prefer}|=1, \lambda=0.3, clip_k=12, \alpha=0.2$. \textbf{Yelp}: $n=100, L=2, d_{model}=64, h=8, d=0.5, L2=0.0, \mathcal{M}=20, \mathcal{K}=9, |\mathcal{T}^{(u)}{pseudo}|=1, |\mathcal{F}^{(u)}{prefer}|=1, \lambda=0.6, clip_k=6, \alpha=0.7$. \textbf{Tenrec}: $n=100, L=2, d_{model}=32, h=2, d=0.5, L2=0.0, \mathcal{M}=18, \mathcal{K}=17, |\mathcal{T}^{(u)}{pseudo}|=1, |\mathcal{F}^{(u)}{prefer}|=1, \lambda=0.3, clip_k=12, \alpha=0.2$.
}
For all datasets except Tenrec, our models are trained for 200 epochs to ensure model convergence.
For model evaluation, we adopted the widely endorsed \emph{leave-one-out} strategy \cite{kang2018self,sun2019bert4rec,liu2021augmenting}, reserving the last item of each user sequence for testing, the penultimate for validation, and the rest for training. Additionally, in alignment with established protocols in \cite{kang2018self,sun2019bert4rec,liu2021augmenting}, we juxtaposed each test item with 100 non-interacted items as negative samples, \revise{excluding Table \ref{tab:main} and Table \ref{tab:large_scale}, which report the ranking results over the entire item set.}
We utilized three evaluation metrics-\emph{Recall (HR)}, \emph{Normalized Discounted Cumulative Gain (NDCG)}, and \emph{Mean Reciprocal Rank (MRR)}-with the results presented being the average values across 10 random seeds.

\begin{table}[t]
  \centering
  \caption{The detailed statistics of dataset \round{after preprocessing}. `\# Inter.' denotes the number of interactions. '\# Avg. Length$_\mathcal{U}$' and `\# Avg. Length$_\mathcal{I}$' denotes average actions of Users/Items. The sparsity levels of datasets equal $1 - \text{density}$.}
  \resizebox{0.9\linewidth}{!}{
    \begin{tabular}{l|cccccc}
    \hline
    Properties & Beauty & Phones & \round{Yelp} & \round{Sports} & \round{TenRec} \\ 
    \hline
    \hline
    \# Users & 22,363 & 27,879 & \round{30,431} & \round{35,598} & \round{983,642} \\ 
    \# Items & 12,101 & 10,429 & \round{20,033} & \round{18,357} & \round{410,112} \\ 
    \# Avg. Length$_\mathcal{U}$ & 8.88 & 6.97 & \round{10.40} & \round{8.32} & \round{37.56} \\ 
    \# Avg. Length$_\mathcal{I}$ & 16.40 & 18.64 & \round{15.79} & \round{16.14} & \round{90.09} \\ 
    \# Interactions & 198,502 & 194,439 & \round{316,354} & \round{296,337} & \round{36,947,008} \\ 
    Density & 0.07\% & 0.06\% & \round{0.05\%} & \round{0.05\%} & \round{0.01\%} \\ 
    \hline
    \end{tabular}
    }
  \label{tab:stat}
  \vspace{-5mm}
\end{table}

\begin{table*}[h]
\centering
\caption{Comparison of recommendation performance with baseline models in ranking results over the whole item set. \textbf{Bold} scores are the best in each row, while \underline{underlined} scores are the second best among baseline models, excluding the variants of ours. $*$, $\dagger$, $\mathsection$, and $\mathparagraph$ denote Transformer-based, data augmentation-based, GNN-based, and contrastive learning-based SR models, respectively. The subscript ``$\text{w/o PT}$'' indicates models without pre-training, meaning the model is fine-tuned from scratch. 
} 
\vspace{2mm}
\label{tab:main}
    \resizebox{0.85\linewidth}{!}{
    \begin{tabular}{c|l|cc|c|c|cc|cc|c}
    \hline
         \multirow{2}*{Dataset} & \multirow{2}*{Metric} & \multicolumn{2}{c|}{$*$} & $\mathsection$ & \multicolumn{1}{c|}{$*$, $\dagger$, and $\mathparagraph$} & \multicolumn{4}{c|}{$*$ and $\dagger$} & \multirow{2}*{Improv.}\\
         \cline{3-10}
         ~ & & SASRec$^\ddagger$ & BERT4Rec$^\ddagger$ & SRGNN & CL4SRec  & $\text{ASReP}_\text{w/o PT}$ & ASReP &  $\text{BARec}_\text{w/o PT}$ & BARec & ~ \\
         \hline
         \hline
         \multirow{4}*{Beauty}&Recall@5& 0.0288  & 0.0216  & 0.0328  & 0.0335  & 0.0251  & \underline{0.0343}  & 0.0295  & \textbf{0.0400} & +16.62\%\\
         &Recall@10& 0.0505  & 0.0377  & 0.0521  & 0.0561   & 0.0471  & \underline{0.0582}  &  0.0521  & \textbf{0.0640} & +9.97\%\\
         &NDCG@5& 0.0151  & 0.0139  & \underline{0.0225}  & 0.0200   & 0.0146  & 0.0213  & 0.0179  & \textbf{0.0256} & +13.78\% \\
         &NDCG@10& 0.0221  & 0.0190  & 0.0286  & 0.0273   & 0.0217  & \underline{0.0289}  & 0.0258  & \textbf{0.0333} & +15.22\%\\
         \hline
         \multirow{4}*{Phones}&Recall@5& 0.0367   & 0.0337   & 0.0432  & 0.0483   & 0.0477   & \underline{0.0527}  & {0.0657}  & \textbf{0.0731} & +38.71\%\\
         &Recall@10& 0.0621   & 0.0524   & 0.0668  & 0.0756   & 0.0728  & \underline{0.0827}  & {0.0927}  & \textbf{0.1045} & +26.36\%\\
         &NDCG@5& 0.0199  & 0.0217    & 0.0285   & 0.0300    & 0.0306   & \underline{0.0348}   & {0.0465}  & \textbf{0.0523} & +50.29\%\\
         &NDCG@10& 0.0281   & 0.0277   & 0.0360  & 0.0388   & 0.0386  & \underline{0.0445}  & {0.0554}  & \textbf{0.0625} & +40.45\%\\
         \hline
         \multirow{4}*{Yelp}&Recall@5& 0.0342  & 0.0147  & 0.0257  & \underline{0.0380}   & 0.0180  & 0.0249  & 0.0359  & \textbf{0.0443} & +16.58\% \\
         &Recall@10& 0.0466  & 0.0256  & 0.0413  & \underline{0.0498}   & 0.0305  & 0.0408  & 0.0446  & \textbf{0.0528} & +6.02\%\\
         &NDCG@5& 0.0268  & 0.0093  & 0.0170  & \underline{0.0291}   & 0.0114  & 0.0161  & 0.0306  & \textbf{0.0367} & +26.12\%\\
         &NDCG@10& 0.0308  & 0.0127  & 0.0220  & \underline{0.0329}   & 0.0153  & 0.0212  & 0.0326  & \textbf{0.0395} & +20.06\%\\
         \hline
         \multirow{4}*{Sports} & Recall@5& 0.0149  & 0.0103  & 0.0149  & \underline{0.0217}   & 0.0122  & 0.0188  & 0.0186  & \textbf{0.0272} & +25.35\%\\
         &Recall@10& 0.0262   & 0.0176   & 0.0250  & \underline{0.0355}   & 0.0224  & 0.0312  & 0.0296  & \textbf{0.0399} & +12.39\% \\
         &NDCG@5& 0.0077  & 0.0063   & 0.0095  & \underline{0.0133}   & 0.0076  & 0.0115  & 0.0130  & \textbf{0.0185} & +39.10\%\\
         &NDCG@10& 0.0114  & 0.0086   & 0.0127  & \underline{0.0178}  & 0.0108  & 0.0155  & 0.0167  & \textbf{0.0226} & +26.97\% \\
         \hline
    \end{tabular}
}
\end{table*}

\begin{table*}
\centering
    \caption{Comparison of recommendation performance on very short sequences ($L \le 3$) and very long sequences ($20 < L \le 50$) with baseline models in numerical values. 
    \textbf{Bold} scores are the best in each column, while \underline{underlined} scores are the second best.}
    \label{tab:short}
    \resizebox{\linewidth}{!}{
    \begin{tabular}{c|c|cccccc|cccccc}
        \hline 
        \multirow{2}*{Dataset} & \multirow{2}*{Model} &  \multicolumn{6}{c|}{Short Sequences ($L \le 3$)} & \multicolumn{6}{c}{Long Sequences ($20 < L \le 50$)} \\
        \cline{3-14}
        ~ & ~ & Recall@1 & Recall@5 & Recall@10 & NDCG@5 & NDCG@10 & MRR & Recall@1 & Recall@5 & Recall@10 & NDCG@5 & NDCG@10 & MRR \\ 
        \hline
        \hline
        \multirow{5}*{Beauty} & SASRec$^a$ & 0.1577 & 0.3626 & 0.4667 &	0.2646 & 0.2981 & 0.2634 & 0.3254 & 0.5882 & 0.6445 & 0.4661 & 0.4844 & 0.4439 \\
         ~ & ASReP$^b$  & \underline{0.2177} & \underline{0.4623} &	\underline{0.5755} & \underline{0.3459} & \underline{0.3827} & \underline{0.3376} & \underline{0.3629} & \underline{0.5932} &	\underline{0.6733} & \underline{0.4877} & \underline{0.5138} & \underline{0.4731} \\
         & BARec & \textbf{0.2485} & \textbf{0.5127} & \textbf{0.6378} & \textbf{0.3869} & \textbf{0.4275} & \textbf{0.3770} & \textbf{0.3954} & \textbf{0.6270} & \textbf{0.7071} & \textbf{0.4929} & \textbf{0.5189} & \textbf{0.4789} \\
        & Improv. a & +57.58\% & +41.40\% & +36.67\% & +46.22\% & +43.41\% & +43.13\% & +21.51\% & +6.60\% & +9.71\% & +5.75\% & +7.12\% & +5.63\% \\
        & Improv. b & +14.15\% &	+10.90\% & +10.83\%	& +11.85\% &	+11.71\%	& +11.67\% & +8.96\% & +5.70\% & +5.02\%	& +1.07\% &	+0.99\%	& +1.23\% \\
        \hline
        \multirow{5}*{Phones} & SASRec$^a$ & 0.1911 & 0.4628 & 0.5894 & 0.3331 & 0.3741 & 0.3224 & \underline{0.2500} & \underline{0.5613} & 0.6650 & \underline{0.4110} & \underline{0.4449} & \underline{0.3877} \\
        & ASReP$^b$ & \underline{0.2508} & \underline{0.5440} & \underline{0.6754} & \underline{0.4046} & \underline{0.4472} & \underline{0.3890} & 0.2169 & 0.5471 & \underline{0.6886} & 0.3964 & 0.4420 & 0.3761 \\ 
        & BARec & \textbf{0.2977} & \textbf{0.5955} & \textbf{0.7230} & \textbf{0.4535} & \textbf{0.4950} & \textbf{0.4363} & \textbf{0.2642} & \textbf{0.6226} & \textbf{0.7453} & \textbf{0.4534} & \textbf{0.4935} & \textbf{0.4232} \\
        & Improv. a & +55.78\% & +28.67\% & +22.67\% & +36.15\% & +32.32\%	& +35.33\% & +5.68\% & +10.92\% & +12.08\% & +10.32\% & +10.92\%	& +9.16\% \\
        & Improv. b & +18.70\% & +9.47\% & +7.05\% & +12.09\% & +10.69\% & +12.16\% & +21.81\% & +13.80\% & +8.23\% & +14.38\% & +11.65\% & +12.52\% \\
        \hline
    \end{tabular}
    }
\end{table*}

\subsection{Overall Performance Comparison (RQ1)}

Table \ref{tab:main} delineates a comparative analysis of average performance across all sequence lengths. 
\round{The improvement over the best-performing baseline methods is statistically significant (sign test, $p < 0.05$).}
Key insights from this analysis are summarized as follows:
\textcircled{1} Transformer-based architectures without data augmentation, specifically SASRec and BERT4Rec, consistently and significantly underperform the data augmentation model ASReP across all evaluated datasets and metrics, a difference that becomes even more pronounced when compared to BARec with data augmentation pre-training.
Specifically, BARec exhibits a marked improvement over SASRec (BERT4Rec), with an average relative enhancement of 62.54\% (141.89\%) in Recall@5 and 102.39\% (178.36\%) in NDCG@5. 
These results strongly underscore the effectiveness of our data augmentation pre-training approach.
\textcircled{2} BARec outperforms ASReP, achieving an average gain of 44.48\% in Recall@5 and 64.83\% in NDCG@5 across all datasets. Unlike ASReP, which solely employs reverse forward pre-training, BARec benefits from a bidirectional temporal pre-training strategy. 
Moreover, BARec also outperforms GNN-based SRGNN and contrastive learning-based CL4SRec, establishing the state-of-the-art performance.
\textcircled{3} Retrained variants, namely $\text{ASReP}_\text{w/o PT}$ and $\text{BARec}_\text{w/o PT}$, exhibit reduced performance compared to their non-retrained counterparts. This observation aligns with the anticipated advantage of pre-training, which imparts a richer semantic understanding of items, particularly beneficial in resource-scarce recommendation contexts. 
Remarkably, $\text{BARec}_\text{w/o PT}$ surpasses ASReP in all metrics on both the Phones and Yelp datasets, a testament to the superior quality of the pseudo-prior items generated by our approach, which in turn provides more informative context (refer to Sec. \ref{sec:pre-training}). 

\subsection{Performance on Short and Long Sequences (RQ2)} 
The evaluation of very short sequences (with $L \le 3$) is illustrated in the left side of Table \ref{tab:short}. For these sequence lengths, BARec demonstrates superior performance over leading models SASRec and ASReP in the Recall@\{1, 5, 10\} metrics, yielding average improvements of 40.46\% and 11.85\%, respectively.  
These findings corroborate the hypothesis that strategic data augmentation for sequences can significantly enhance the performance on short sequences, thereby offering a viable solution to alleviate the challenges associated with the cold-start problem in recommendation systems. 
We further examine our model's capability to harness the extensive contextual information inherent in long sequences for enhanced performance. 
As demonstrated in the right side of Table \ref{tab:short}, for sequences of length $20 < L \le 50$, BARec exhibits superior performance over SASRec and ASReP in the Recall@\{1, 5, 10\} metrics, with average improvements of 11.08\% and 10.59\% across all benchmarks, respectively.  
These results indicate that BARec is adept at extracting valuable contextual information from long sequences, thereby enabling a more precise representation of user behavior patterns.

\begin{figure}[t]
\begin{center}
\centering
\subfigure{\includegraphics[width=0.465\linewidth]{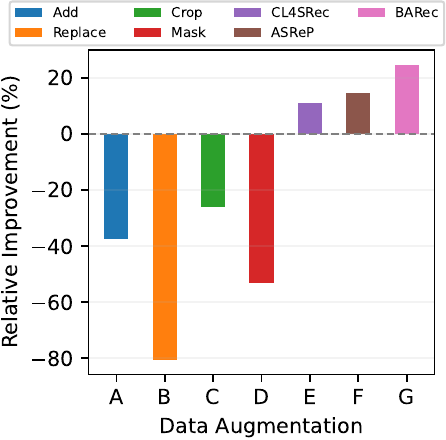}}
\hfill
\subfigure{\includegraphics[width=0.465\linewidth]{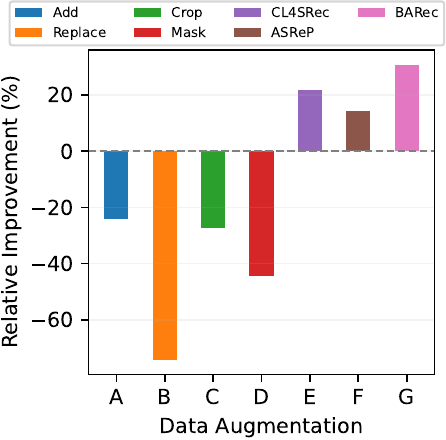}}
\caption{Performance comparison of a variety of data augmentation strategies for sequential recommendation. The relative improvements over baseline SASRec \cite{kang2018self} with Recall@10 metric on Beauty (left side) and Phones (right side) are reported.}
\label{fig:data_aug}    
\end{center}
\vspace{-5mm}
\end{figure}
\subsection{Efficacy of Data Augmentation Strategies (RQ3)} 
SR are particularly sensitive to the temporal order of items and the dynamic preferences of users, where any random perturbation to the sequence can obscure these user preferences. 
To explore the influence of data augmentation strategies on the predictive quality of SR models, we conducted an evaluation of various augmentation strategies.
Our analysis systematically categorizes existing data augmentation methods for SR into two primary groups: random perturbation strategies, encompassing item cropping, item masking, item addition, and item replacement, and generative approaches, which include reverse sequence generation (ASReP), bidirectional generation (BARec), and contrastive learning (CL4SRec). 
These methods were benchmarked against a baseline non-data-augmentation approach (SASRec) and evaluated using Recall@10 on two distinct datasets: Beauty and Phones.
The results of relative improvements, presented in Figure \ref{fig:data_aug}, indicate that the random perturbation strategies generally lead to a decline in SASRec's performance, which can be attributed to the inherent complexities of SR tasks, such as the discrete sequence dynamics, dependencies between items, and the varied lengths of sequences. 
On the contrary, the generative augmentation approaches, ASReP and BARec, demonstrated an average performance enhancement of 14.46\% and 27.76\%, respectively, underscoring the potential benefits of generative data augmentation strategies in this context. 
Moreover, our proposed BARec model exhibits a noteworthy average relative improvement of 13.3\% over ASReP, highlighting its effectiveness in synthesizing pseudo-prior items of high quality.

\begin{figure*}[t]
\begin{center}
\centerline{
\begin{minipage}[b]{0.85\linewidth} 
    \subfigure{\includegraphics[width=0.245\linewidth]{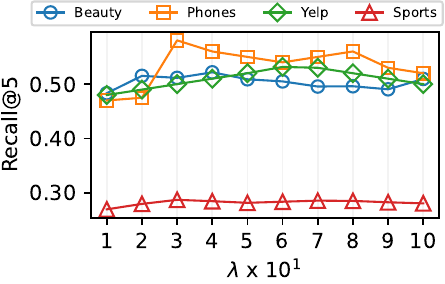}}
    \hfill
    \subfigure{\includegraphics[width=0.245\linewidth]{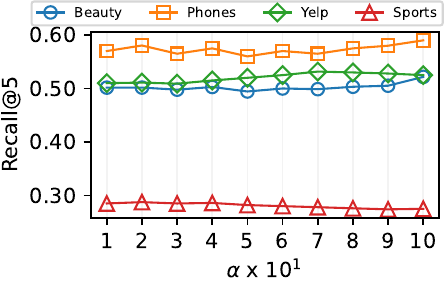}}
    \hfill
    \subfigure{\includegraphics[width=0.245\linewidth]{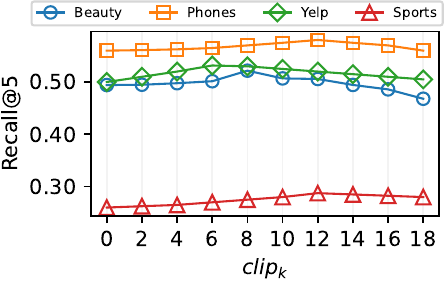}}
    \hfill
    \subfigure{\includegraphics[width=0.245\linewidth]{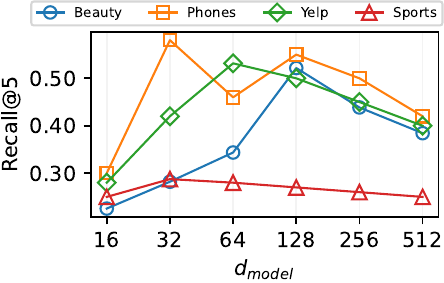}}
    \hfill
    \subfigure{\includegraphics[width=0.245\linewidth]{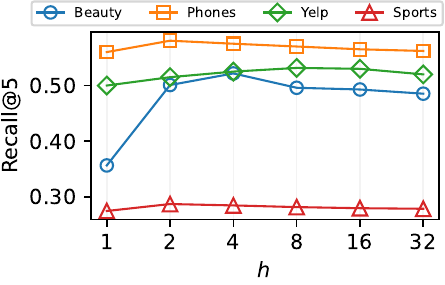}}
    \hfill
    \subfigure{\includegraphics[width=0.245\linewidth]{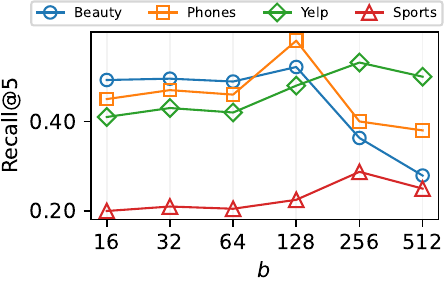}}
    \hfill
    \subfigure{\includegraphics[width=0.245\linewidth]{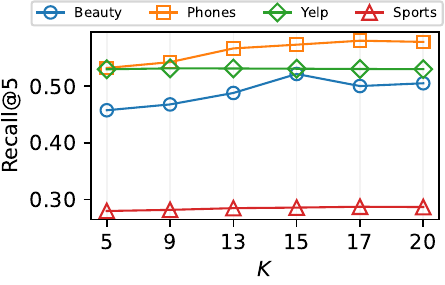}}
    \hfill
    \subfigure{\includegraphics[width=0.245\linewidth]{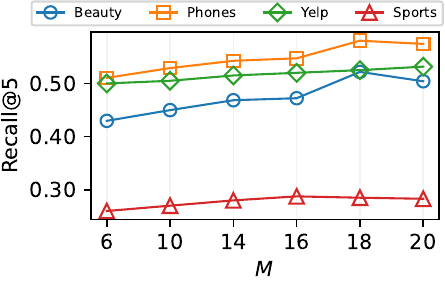}}
\end{minipage}}
\caption{Comprehensive hyper-parameters sensitivity analysis, including trade-off coefficient $\lambda$, balance coefficient $\alpha$, clipped gold labels $clip_k$, model dimension $d_{model}$, attention heads $h$, batch size $b$, pseudo-prior items $\mathcal{K}$, and sequence length threshold $\mathcal{M}$, as applied to the Beauty dataset.}
\label{fig:hyper}    
\end{center}
\vspace{-5mm}
\end{figure*}

\subsection{Ablation Study of Data Augmentation (RQ4)}
To discern the individual contributions of the proposed components, we executed a comprehensive ablation study, the results of which are presented in Table \ref{tab:ablation} for the Beauty and Phones. 
We introduce the notations \textsc{Bi} for bidirectional temporal pre-training, \textsc{AUG} for pseudo-prior item-based data augmentation, \textsc{KL} for model knowledge enhancement, and \textsc{LC} for the gold label clipping operation. 
Our findings are as follows:
1) The results, as depicted in pairs (C)(E) and (H)(J), indicate that \textsc{Bi} enhances performance with an average increment of 7.54\% in NDCG@5. This substantiates the alignment of \textsc{Bi} with the SR task, affirming its efficacy in capturing bidirectional temporal dynamics.
2) The results in pairs (A)(B) and (F)(G) reveal that \textsc{AUG} leads to a substantial performance boost, with an average improvement of 22.19\% in NDCG@5. This underscores the potency of \textsc{AUG} as a data augmentation strategy, preserving user preference continuity in the SR context.
3) As shown by the comparisons in (D)(E) and (I)(J), \textsc{KL} contributes to a performance increase of 4.15\% in NDCG@5 on average. This enhancement can be attributed to the augmented representational power of the model, which, in turn, facilitates more effective fine-tuning. Concurrently, \textsc{LC} addresses the limitations posed by short sequence lengths, thereby offering a partial solution to the information scarcity in such scenarios.

\begin{table}[t]
\caption{Ablation study of the model with({\color{black}\CheckmarkBold}) or without({\color{black}\XSolidBrush}) proposed components. $\bm{+\Delta}$ denotes the average relative improvement in terms of NDCG@5.}
\label{tab:ablation}
\centering
\resizebox{0.8\linewidth}{!}{
\centering
\begin{tabular}{c|c|cccc|cc|c}
\hline
\multirow{2}*{Dataset}& \multirow{2}*{Setting}&\multicolumn{4}{c|}{Components} & \multicolumn{3}{c}{Results} \\
\cline{3-9}
& &\textsc{Bi} & \textsc{AUG} & \textsc{KL} & \textsc{LC} & Recall@5 & NDCG@5 & $+\Delta$\\ 
\hline
\hline
\multirow{5}*{Beauty}& (A) &\color{black}\XSolidBrush& \color{black}\XSolidBrush & \color{black}\XSolidBrush & \color{black}\XSolidBrush & 0.3963 & 0.2949 & - \\ 
&(B) &\color{black}\XSolidBrush& \color{black}\CheckmarkBold & \color{black}\XSolidBrush & \color{black}\XSolidBrush & 0.4684 & 0.3547 & +20.28\%\\ 
&(C) &\color{black}\XSolidBrush& \color{black}\CheckmarkBold & \color{black}\CheckmarkBold & \color{black}\CheckmarkBold & 0.4810 & 0.3573 & +21.16\%\\ 
&(D) &\color{black}\CheckmarkBold& \color{black}\CheckmarkBold & \color{black}\XSolidBrush & \color{black}\CheckmarkBold & 0.4868 & 0.3671 & +24.48\% \\ 
&(E) &\color{black}\CheckmarkBold& \color{black}\CheckmarkBold & \color{black}\CheckmarkBold & \color{black}\CheckmarkBold & \textbf{0.5217} & \textbf{0.3892}  & \textbf{+31.98\%} \\ 
\hline
\multirow{5}*{Phones}&(F) &\color{black}\XSolidBrush& \color{black}\XSolidBrush & \color{black}\XSolidBrush & \color{black}\XSolidBrush & 0.4646 & 0.3379  & -\\
&(G) &\color{black}\XSolidBrush& \color{black}\CheckmarkBold & \color{black}\XSolidBrush & \color{black}\XSolidBrush & 0.5573 & 0.4193 & +24.09\%\\ 
&(H) &\color{black}\XSolidBrush& \color{black}\CheckmarkBold & \color{black}\CheckmarkBold & \color{black}\CheckmarkBold & 0.5546 &  0.4118 & +21.87\%\\ 
&(I) &\color{black}\CheckmarkBold& \color{black}\CheckmarkBold & \color{black}\XSolidBrush & \color{black}\CheckmarkBold & 0.5635 & 0.4274 & +26.49\% \\ 
&(J) &\color{black}\CheckmarkBold& \color{black}\CheckmarkBold & \color{black}\CheckmarkBold & \color{black}\CheckmarkBold & \textbf{0.5804} & \textbf{0.4371} & \textbf{+29.36\%} \\
\hline
\end{tabular}}
\vspace{-5mm}
\end{table}

\round{
\subsection{Hyper-parameters Sensitivity Study (RQ5)}
This section explores the impact of hyper-parameters, including $\lambda$, $\alpha$, $clip_k$, $d_{model}$, $h$, $b$, $\mathcal{K}$, and $\mathcal{M}$, on our model's performance, with results depicted in Fig. \ref{fig:hyper}. 
Notably, we observe a unimodal relationship between model performance and the aforementioned hyperparameters on Beauty, demonstrating initial improvement followed by a decline as the parameters increase—except for $\alpha$, which exhibits a negligible marginal effect. 
In addition, we scrutinize the influence of pivotal hyperparameters $K$ and $M$, corresponding to the count of pseudo-prior items and the threshold for short sequences, respectively. 
For each metric, we isolate the variation of a single hyperparameter while maintaining others at their optimal levels. Optimal performance is achieved on all datasets when $K$ and $M$ are within the 15 to 18 range. 
These findings affirm the critical nature of precise hyperparameter tuning, with the guiding principle being to avoid overly restrictive values (leading to inadequate contextual information) and excessively high ones (introducing inefficiencies).
}

\revise{
\subsection{Large-scale Recommendation Scenarios (RQ6)}
Current benchmark datasets for recommender systems (RS) are typically constrained by their small scale or by the narrow scope of user feedback \cite{yuan2022tenrec,gao2022kuairec}. 
To more accurately represent the complexity of real-world recommendation scenarios, we employ the large-scale Tenrec\footnote{\scriptsize{\url{https://github.com/yuangh-x/2022-NIPS-Tenrec}}} dataset \cite{yuan2022tenrec}, a comprehensive and publicly available dataset curated specifically for RS. 
It encompasses a wide range of user feedback across four unique recommendation scenarios. Our study concentrates on the QK-video category. 
Given the extensive volume of QK-video data and the computational limitations we faced, our model training was confined to a single epoch.
The results, alongside corresponding leaderboard\footnote{\scriptsize{\url{https://tenrec0.github.io/\#leaderboard}}} results, are detailed in Table \ref{tab:large_scale}.
The empirical evidence clearly demonstrates that models incorporating data augmentation techniques (ASReP and BARec) significantly surpass traditional non-data-augmentation methodologies (SASRec and BERT4Rec) in performance. 
This advantage underscores the effectiveness of leveraging pseudo-prior item data augmentation as a strategic approach to mitigate the prevalent issue of data scarcity in RS.
}

\begin{table}[t]
  \centering
  \caption{Comparison of recommendation performance with baseline models in ranking results over the entire item set on the established large-scale public Tenrec \cite{yuan2022tenrec} benchmark. The results are evaluated on QK-video-1M.}
  \resizebox{\linewidth}{!}{
    \begin{tabular}{l|ccccc|c}
    \hline
    \multirow{2}*{Model} & \multicolumn{5}{c|}{Our Results}& Leaderboard  \\
    \cline{2-7}
    ~ & Recall@5 & Recall@10 & NDCG@5 & NDCG@10 & NDCG@20 & NDCG@20  \\
    \hline
    \hline
    SASRec & 0.0114 & 0.0200 & 0.0070 & 0.0097 & 0.0133 & 0.0194 \\
    BERT4Rec & 0.0102 & 0.0178 & 0.0061 & 0.0086 & 0.0116 &  0.0185 \\
    ASReP & 0.0216 & 0.0378 & 0.0133 & 0.0189 &0.0259 & - \\
    \hline
    BARec & 0.0287 & 0.0508 & 0.0174 & 0.0245 & 0.0334 & - \\
    \hline
    \end{tabular}
    }
  \label{tab:large_scale}
  \vspace{-5mm}
\end{table}

\round{
\subsection{Case Study}
\begin{figure}[t]
\centering
\includegraphics[width=\linewidth]{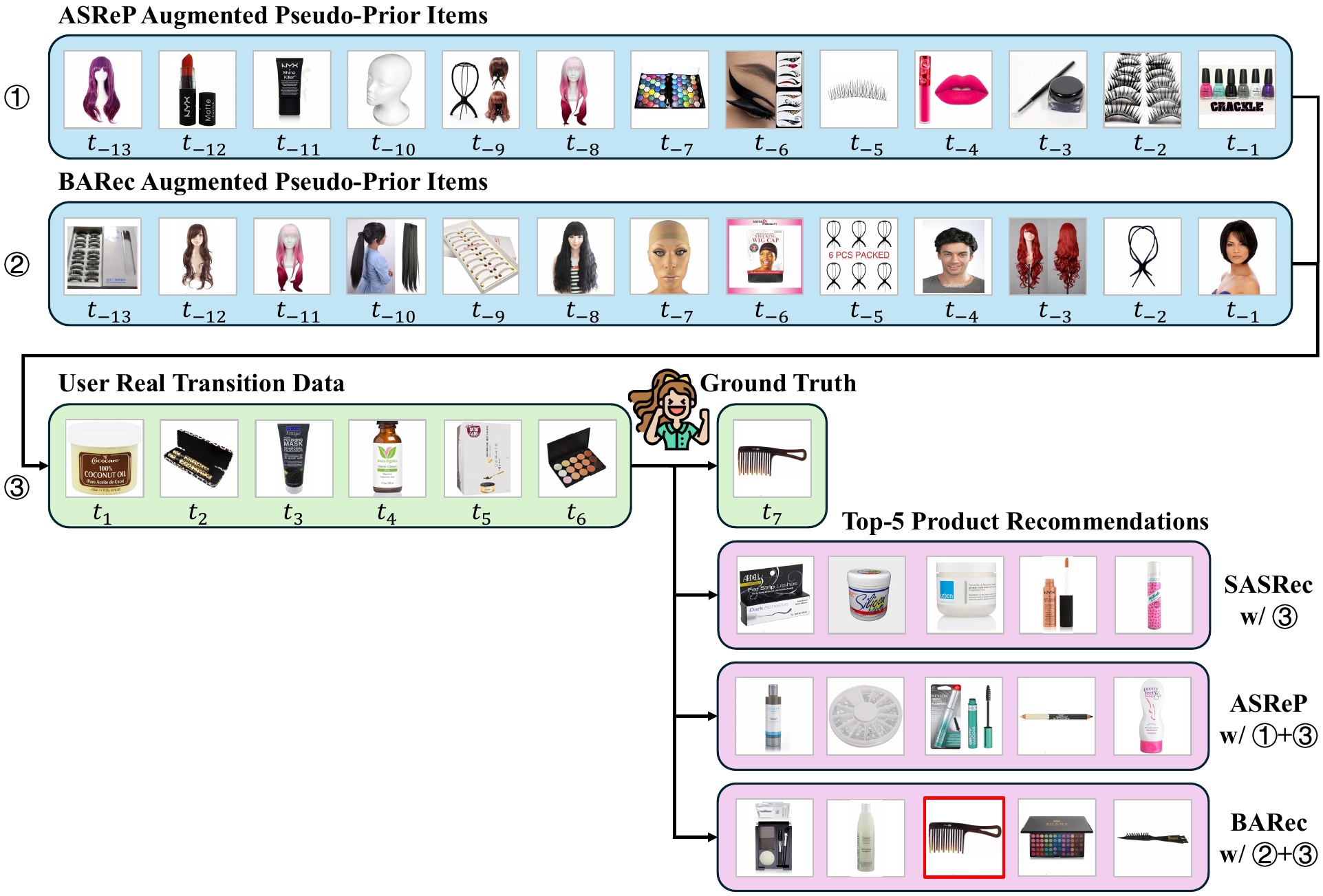}
\caption{A case study on a randomly selected user from Amazon's product data within the Beauty category. The ASReP and BARec methods employ data augmentation strategies to enrich the context of short sequences by generating pseudo-prior items. The results of the top-5 product recommendations using the SASRec, ASReP, and BARec models indicate that our BARec model more accurately captures user preference patterns.
}
\label{fig:case_study}
\end{figure}
To better demonstrate why our proposed BARec achieves notable performance improvements, 
we present a case study on a randomly selected user from Amazon's Beauty category product data, as illustrated in Figure \ref{fig:case_study} of this response. 
Our observations reveal that, in comparison to ASReP data augmentation strategy, our BARec method generates a greater number of items that are closely related to hair care. 
This demonstrates that BARec effectively preserves the temporal correlations within the generated pseudo-prior items, aligning seamlessly with the user's actual transition data.
The results of the top-5 product recommendations using the SASRec, ASReP, and BARec models highlight the following insights:
\begin{itemize}
    \item By employing data augmentation strategies to enhance the context of short sequences through the generation of pseudo-prior items, both the ASReP and BARec models are better equipped to capture user preferences. This enables more accurate product recommendations based on the user's historical sequential data.
    \item Compared to ASReP, our BARec model successfully identifies user's actual preferences and recommends the ground truth product, ``Mebco Tortoise Shower Detangler'', by leveraging the user preference-preserved augmented pseudo-prior hair-care-related items.
\end{itemize}
}

\section{Conclusion}
In this work, we introduce BARec, an innovative bidirectional temporal data augmentation with pre-training methodology coupled with knowledge-enhanced fine-tuning. This method adeptly synthesizes pseudo-prior items that preserve user preferences and concurrently discerns deeper item semantic interrelations. Rigorous evaluations conducted on five benchmark datasets affirm the superior performance of BARec. Through theoretical analysis, we gain valuable perspectives on the model's inferential capabilities and its interpretative power.
For future studies, we will explore the integration of more sophisticated generative architectures, such as large language models (LLMs), to serve as the foundational framework for our system. 


%



\ifCLASSOPTIONcaptionsoff
  \newpage
\fi



%


\small{
\bibliographystyle{IEEEtran}
\bibliography{ref}

\begin{thebibliography}{10}
\providecommand{\url}[1]{#1}
\csname url@samestyle\endcsname
\providecommand{\newblock}{\relax}
\providecommand{\bibinfo}[2]{#2}
\providecommand{\BIBentrySTDinterwordspacing}{\spaceskip=0pt\relax}
\providecommand{\BIBentryALTinterwordstretchfactor}{4}
\providecommand{\BIBentryALTinterwordspacing}{\spaceskip=\fontdimen2\font plus
\BIBentryALTinterwordstretchfactor\fontdimen3\font minus \fontdimen4\font\relax}
\providecommand{\BIBforeignlanguage}[2]{{%
\expandafter\ifx\csname l@#1\endcsname\relax
\typeout{** WARNING: IEEEtran.bst: No hyphenation pattern has been}%
\typeout{** loaded for the language `#1'. Using the pattern for}%
\typeout{** the default language instead.}%
\else
\language=\csname l@#1\endcsname
\fi
#2}}
\providecommand{\BIBdecl}{\relax}
\BIBdecl

\bibitem{liu2016context}
Q.~Liu, S.~Wu, D.~Wang, Z.~Li, and L.~Wang, ``Context-aware sequential recommendation,'' in \emph{2016 IEEE 16th International Conference on Data Mining (ICDM)}.\hskip 1em plus 0.5em minus 0.4em\relax IEEE, 2016, pp. 1053--1058.

\bibitem{liu2021augmenting}
Z.~Liu, Z.~Fan, Y.~Wang, and P.~S. Yu, ``Augmenting sequential recommendation with pseudo-prior items via reversely pre-training transformer,'' in \emph{Proceedings of the 44th international ACM SIGIR conference on Research and development in information retrieval}, 2021, pp. 1608--1612.

\bibitem{yu2016dynamic}
F.~Yu, Q.~Liu, S.~Wu, L.~Wang, and T.~Tan, ``A dynamic recurrent model for next basket recommendation,'' in \emph{Proceedings of the 39th International ACM SIGIR conference on Research and Development in Information Retrieval}, 2016, pp. 729--732.

\bibitem{ma2019hierarchical}
C.~Ma, P.~Kang, and X.~Liu, ``Hierarchical gating networks for sequential recommendation,'' in \emph{Proceedings of the 25th ACM SIGKDD international conference on knowledge discovery \& data mining}, 2019, pp. 825--833.

\bibitem{kang2018self}
W.-C. Kang and J.~McAuley, ``Self-attentive sequential recommendation,'' in \emph{2018 IEEE International Conference on Data Mining (ICDM)}.\hskip 1em plus 0.5em minus 0.4em\relax IEEE, 2018, pp. 197--206.

\bibitem{jiang2023adamct}
J.~Jiang, P.~Zhang, Y.~Luo, C.~Li, J.~B. Kim, K.~Zhang, S.~Wang, X.~Xie, and S.~Kim, ``Adamct: adaptive mixture of cnn-transformer for sequential recommendation,'' in \emph{Proceedings of the 32nd ACM International Conference on Information and Knowledge Management}, 2023, pp. 976--986.

\bibitem{sun2019bert4rec}
F.~Sun, J.~Liu, J.~Wu, C.~Pei, X.~Lin, W.~Ou, and P.~Jiang, ``Bert4rec: Sequential recommendation with bidirectional encoder representations from transformer,'' in \emph{Proceedings of the 28th ACM international conference on information and knowledge management}, 2019, pp. 1441--1450.

\bibitem{wang2021sequential}
J.~Wang, K.~Ding, and J.~Caverlee, ``Sequential recommendation for cold-start users with meta transitional learning,'' in \emph{Proceedings of the 44th International ACM SIGIR Conference on Research and Development in Information Retrieval}, 2021, pp. 1783--1787.

\bibitem{liu2021contrastive}
Z.~Liu, Y.~Chen, J.~Li, P.~S. Yu, J.~McAuley, and C.~Xiong, ``Contrastive self-supervised sequential recommendation with robust augmentation,'' \emph{arXiv preprint arXiv:2108.06479}, 2021.

\bibitem{qiu2022contrastive}
R.~Qiu, Z.~Huang, H.~Yin, and Z.~Wang, ``Contrastive learning for representation degeneration problem in sequential recommendation,'' in \emph{Proceedings of the fifteenth ACM international conference on web search and data mining}, 2022, pp. 813--823.

\bibitem{vasile2016meta}
F.~Vasile, E.~Smirnova, and A.~Conneau, ``Meta-prod2vec: Product embeddings using side-information for recommendation,'' in \emph{Proceedings of the 10th ACM conference on recommender systems}, 2016, pp. 225--232.

\bibitem{chen2018collective}
Y.~Chen and M.~de~Rijke, ``A collective variational autoencoder for top-n recommendation with side information,'' in \emph{Proceedings of the 3rd workshop on deep learning for recommender systems}, 2018, pp. 3--9.

\bibitem{liu2021noninvasive}
C.~Liu, X.~Li, G.~Cai, Z.~Dong, H.~Zhu, and L.~Shang, ``Noninvasive self-attention for side information fusion in sequential recommendation,'' in \emph{Proceedings of the AAAI Conference on Artificial Intelligence}, vol.~35, no.~5, 2021, pp. 4249--4256.

\bibitem{deldjoo2020recommender}
Y.~Deldjoo, M.~Schedl, P.~Cremonesi, and G.~Pasi, ``Recommender systems leveraging multimedia content,'' \emph{ACM Computing Surveys (CSUR)}, vol.~53, no.~5, pp. 1--38, 2020.

\bibitem{kulkarni2020context}
S.~Kulkarni and S.~F. Rodd, ``Context aware recommendation systems: A review of the state of the art techniques,'' \emph{Computer Science Review}, vol.~37, p. 100255, 2020.

\bibitem{milano2020recommender}
S.~Milano, M.~Taddeo, and L.~Floridi, ``Recommender systems and their ethical challenges,'' \emph{Ai \& Society}, vol.~35, pp. 957--967, 2020.

\bibitem{li2017neural}
J.~Li, P.~Ren, Z.~Chen, Z.~Ren, T.~Lian, and J.~Ma, ``Neural attentive session-based recommendation,'' in \emph{Proceedings of the 2017 ACM on Conference on Information and Knowledge Management}, 2017, pp. 1419--1428.

\bibitem{yuan2020future}
F.~Yuan, X.~He, H.~Jiang, G.~Guo, J.~Xiong, Z.~Xu, and Y.~Xiong, ``Future data helps training: Modeling future contexts for session-based recommendation,'' in \emph{Proceedings of The Web Conference 2020}, 2020, pp. 303--313.

\bibitem{wang2021counterfactual}
Z.~Wang, J.~Zhang, H.~Xu, X.~Chen, Y.~Zhang, W.~X. Zhao, and J.-R. Wen, ``Counterfactual data-augmented sequential recommendation,'' in \emph{Proceedings of the 44th International ACM SIGIR Conference on Research and Development in Information Retrieval}, 2021, pp. 347--356.

\bibitem{xie2022contrastive}
X.~Xie, F.~Sun, Z.~Liu, S.~Wu, J.~Gao, J.~Zhang, B.~Ding, and B.~Cui, ``Contrastive learning for sequential recommendation,'' in \emph{2022 IEEE 38th international conference on data engineering (ICDE)}.\hskip 1em plus 0.5em minus 0.4em\relax IEEE, 2022, pp. 1259--1273.

\bibitem{zhou2020s3}
K.~Zhou, H.~Wang, W.~X. Zhao, Y.~Zhu, S.~Wang, F.~Zhang, Z.~Wang, and J.-R. Wen, ``S3-rec: Self-supervised learning for sequential recommendation with mutual information maximization,'' in \emph{Proceedings of the 29th ACM International Conference on Information \& Knowledge Management}, 2020, pp. 1893--1902.

\bibitem{duorec}
R.~Qiu, Z.~Huang, H.~Yin, and Z.~Wang, ``Contrastive learning for representation degeneration problem in sequential recommendation,'' in \emph{{WSDM}}.\hskip 1em plus 0.5em minus 0.4em\relax {ACM}, 2022, pp. 813--823.

\bibitem{zhou2023equivariant}
P.~Zhou, J.~Gao, Y.~Xie, Q.~Ye, Y.~Hua, J.~Kim, S.~Wang, and S.~Kim, ``Equivariant contrastive learning for sequential recommendation,'' in \emph{Proceedings of the 17th ACM Conference on Recommender Systems}, 2023, pp. 129--140.

\bibitem{kenton2019bert}
J.~D. M.-W.~C. Kenton and L.~K. Toutanova, ``Bert: Pre-training of deep bidirectional transformers for language understanding,'' in \emph{Proceedings of NAACL-HLT}, 2019, pp. 4171--4186.

\bibitem{radford2018improving}
A.~Radford, K.~Narasimhan, T.~Salimans, I.~Sutskever \emph{et~al.}, ``Improving language understanding by generative pre-training,'' 2018.

\bibitem{shorten2019survey}
C.~Shorten and T.~M. Khoshgoftaar, ``A survey on image data augmentation for deep learning,'' \emph{Journal of big data}, vol.~6, no.~1, pp. 1--48, 2019.

\bibitem{tan2016improved}
Y.~K. Tan, X.~Xu, and Y.~Liu, ``Improved recurrent neural networks for session-based recommendations,'' in \emph{Proceedings of the 1st workshop on deep learning for recommender systems}, 2016, pp. 17--22.

\bibitem{wang2019enhancing}
Q.~Wang, H.~Yin, H.~Wang, Q.~V.~H. Nguyen, Z.~Huang, and L.~Cui, ``Enhancing collaborative filtering with generative augmentation,'' in \emph{Proceedings of the 25th ACM SIGKDD International Conference on Knowledge Discovery \& Data Mining}, 2019, pp. 548--556.

\bibitem{li2019context}
Y.~Li, Y.~Luo, Z.~Zhang, S.~Sadiq, and P.~Cui, ``Context-aware attention-based data augmentation for poi recommendation,'' in \emph{2019 IEEE 35th International Conference on Data Engineering Workshops (ICDEW)}.\hskip 1em plus 0.5em minus 0.4em\relax IEEE, 2019, pp. 177--184.

\bibitem{wang2023diffusion}
W.~Wang, Y.~Xu, F.~Feng, X.~Lin, X.~He, and T.-S. Chua, ``Diffusion recommender model,'' in \emph{Proceedings of the 46th International ACM SIGIR Conference on Research and Development in Information Retrieval}, 2023, pp. 832--841.

\bibitem{liu2023diffusion}
Q.~Liu, F.~Yan, X.~Zhao, Z.~Du, H.~Guo, R.~Tang, and F.~Tian, ``Diffusion augmentation for sequential recommendation,'' in \emph{Proceedings of the 32nd ACM International Conference on Information and Knowledge Management}, 2023, pp. 1576--1586.

\bibitem{yang2024generate}
Z.~Yang, J.~Wu, Z.~Wang, X.~Wang, Y.~Yuan, and X.~He, ``Generate what you prefer: Reshaping sequential recommendation via guided diffusion,'' \emph{Advances in Neural Information Processing Systems}, vol.~36, 2024.

\bibitem{saveski2014item}
M.~Saveski and A.~Mantrach, ``Item cold-start recommendations: learning local collective embeddings,'' in \emph{Proceedings of the 8th ACM Conference on Recommender systems}, 2014, pp. 89--96.

\bibitem{cai2023user}
D.~Cai, S.~Qian, Q.~Fang, J.~Hu, and C.~Xu, ``User cold-start recommendation via inductive heterogeneous graph neural network,'' \emph{ACM Transactions on Information Systems}, vol.~41, no.~3, pp. 1--27, 2023.

\bibitem{wei2021contrastive}
Y.~Wei, X.~Wang, Q.~Li, L.~Nie, Y.~Li, X.~Li, and T.-S. Chua, ``Contrastive learning for cold-start recommendation,'' in \emph{Proceedings of the 29th ACM International Conference on Multimedia}, 2021, pp. 5382--5390.

\bibitem{du2022socially}
J.~Du, Z.~Ye, L.~Yao, B.~Guo, and Z.~Yu, ``Socially-aware dual contrastive learning for cold-start recommendation,'' in \emph{Proceedings of the 45th International ACM SIGIR Conference on Research and Development in Information Retrieval}, 2022, pp. 1927--1932.

\bibitem{zhao2020catn}
C.~Zhao, C.~Li, R.~Xiao, H.~Deng, and A.~Sun, ``Catn: Cross-domain recommendation for cold-start users via aspect transfer network,'' in \emph{Proceedings of the 43rd International ACM SIGIR Conference on Research and Development in Information Retrieval}, 2020, pp. 229--238.

\bibitem{zhu2021transfer}
Y.~Zhu, K.~Ge, F.~Zhuang, R.~Xie, D.~Xi, X.~Zhang, L.~Lin, and Q.~He, ``Transfer-meta framework for cross-domain recommendation to cold-start users,'' in \emph{Proceedings of the 44th International ACM SIGIR Conference on Research and Development in Information Retrieval}, 2021, pp. 1813--1817.

\bibitem{zheng2021cold}
Y.~Zheng, S.~Liu, Z.~Li, and S.~Wu, ``Cold-start sequential recommendation via meta learner,'' in \emph{Proceedings of the AAAI Conference on Artificial Intelligence}, vol.~35, no.~5, 2021, pp. 4706--4713.

\bibitem{lu2020meta}
Y.~Lu, Y.~Fang, and C.~Shi, ``Meta-learning on heterogeneous information networks for cold-start recommendation,'' in \emph{Proceedings of the 26th ACM SIGKDD International Conference on Knowledge Discovery \& Data Mining}, 2020, pp. 1563--1573.

\bibitem{rendle2010factorizing}
S.~Rendle, C.~Freudenthaler, and L.~Schmidt-Thieme, ``Factorizing personalized markov chains for next-basket recommendation,'' in \emph{Proceedings of the 19th international conference on World wide web}, 2010, pp. 811--820.

\bibitem{he2016fusing}
R.~He and J.~McAuley, ``Fusing similarity models with markov chains for sparse sequential recommendation,'' in \emph{2016 IEEE 16th International Conference on Data Mining (ICDM)}.\hskip 1em plus 0.5em minus 0.4em\relax IEEE, 2016, pp. 191--200.

\bibitem{hidasi2015session}
B.~Hidasi, A.~Karatzoglou, L.~Baltrunas, and D.~Tikk, ``Session-based recommendations with recurrent neural networks,'' \emph{arXiv preprint arXiv:1511.06939}, 2015.

\bibitem{quadrana2017personalizing}
M.~Quadrana, A.~Karatzoglou, B.~Hidasi, and P.~Cremonesi, ``Personalizing session-based recommendations with hierarchical recurrent neural networks,'' in \emph{Proceedings of the Eleventh ACM Conference on Recommender Systems}, 2017, pp. 130--137.

\bibitem{song2019hierarchical}
K.~Song, M.~Ji, S.~Park, and I.-C. Moon, ``Hierarchical context enabled recurrent neural network for recommendation,'' in \emph{Proceedings of the AAAI Conference on Artificial Intelligence}, vol.~33, no.~01, 2019, pp. 4983--4991.

\bibitem{smirnova2017contextual}
E.~Smirnova and F.~Vasile, ``Contextual sequence modeling for recommendation with recurrent neural networks,'' in \emph{Proceedings of the 2nd workshop on deep learning for recommender systems}, 2017, pp. 2--9.

\bibitem{hidasi2018recurrent}
B.~Hidasi and A.~Karatzoglou, ``Recurrent neural networks with top-k gains for session-based recommendations,'' in \emph{Proceedings of the 27th ACM international conference on information and knowledge management}, 2018, pp. 843--852.

\bibitem{tang2018personalized}
J.~Tang and K.~Wang, ``Personalized top-n sequential recommendation via convolutional sequence embedding,'' in \emph{Proceedings of the Eleventh ACM International Conference on Web Search and Data Mining}, 2018, pp. 565--573.

\bibitem{yuan2019simple}
F.~Yuan, A.~Karatzoglou, I.~Arapakis, J.~M. Jose, and X.~He, ``A simple convolutional generative network for next item recommendation,'' in \emph{Proceedings of the twelfth ACM international conference on web search and data mining}, 2019, pp. 582--590.

\bibitem{kang2019recommender}
K.~Kang, J.~Park, W.~Kim, H.~Choe, and J.~Choo, ``Recommender system using sequential and global preference via attention mechanism and topic modeling,'' in \emph{Proceedings of the 28th ACM international conference on information and knowledge management}, 2019, pp. 1543--1552.

\bibitem{luo2021stan}
Y.~Luo, Q.~Liu, and Z.~Liu, ``Stan: Spatio-temporal attention network for next location recommendation,'' in \emph{Proceedings of the Web Conference 2021}, 2021, pp. 2177--2185.

\bibitem{wu2019session}
S.~Wu, Y.~Tang, Y.~Zhu, L.~Wang, X.~Xie, and T.~Tan, ``Session-based recommendation with graph neural networks,'' in \emph{Proceedings of the AAAI conference on artificial intelligence}, vol.~33, no.~01, 2019, pp. 346--353.

\bibitem{chang2021sequential}
J.~Chang, C.~Gao, Y.~Zheng, Y.~Hui, Y.~Niu, Y.~Song, D.~Jin, and Y.~Li, ``Sequential recommendation with graph neural networks,'' in \emph{Proceedings of the 44th international ACM SIGIR conference on research and development in information retrieval}, 2021, pp. 378--387.

\bibitem{zhang2022dynamic}
M.~Zhang, S.~Wu, X.~Yu, Q.~Liu, and L.~Wang, ``Dynamic graph neural networks for sequential recommendation,'' \emph{IEEE Transactions on Knowledge and Data Engineering}, vol.~35, no.~5, pp. 4741--4753, 2022.

\bibitem{su2023enhancing}
J.~Su, C.~Chen, W.~Liu, F.~Wu, X.~Zheng, and H.~Lyu, ``Enhancing hierarchy-aware graph networks with deep dual clustering for session-based recommendation,'' in \emph{Proceedings of the ACM Web Conference 2023}, 2023, pp. 165--176.

\bibitem{yuan2022multi}
E.~Yuan, W.~Guo, Z.~He, H.~Guo, C.~Liu, and R.~Tang, ``Multi-behavior sequential transformer recommender,'' in \emph{Proceedings of the 45th international ACM SIGIR conference on research and development in information retrieval}, 2022, pp. 1642--1652.

\bibitem{yang2022multi}
Y.~Yang, C.~Huang, L.~Xia, Y.~Liang, Y.~Yu, and C.~Li, ``Multi-behavior hypergraph-enhanced transformer for sequential recommendation,'' in \emph{Proceedings of the 28th ACM SIGKDD conference on knowledge discovery and data mining}, 2022, pp. 2263--2274.

\bibitem{su2023personalized}
J.~Su, C.~Chen, Z.~Lin, X.~Li, W.~Liu, and X.~Zheng, ``Personalized behavior-aware transformer for multi-behavior sequential recommendation,'' in \emph{Proceedings of the 31st ACM International Conference on Multimedia}, 2023, pp. 6321--6331.

\bibitem{ma2019pi}
M.~Ma, P.~Ren, Y.~Lin, Z.~Chen, J.~Ma, and M.~d. Rijke, ``$\pi$-net: A parallel information-sharing network for shared-account cross-domain sequential recommendations,'' in \emph{Proceedings of the 42nd international ACM SIGIR conference on research and development in information retrieval}, 2019, pp. 685--694.

\bibitem{zheng2022ddghm}
X.~Zheng, J.~Su, W.~Liu, and C.~Chen, ``Ddghm: Dual dynamic graph with hybrid metric training for cross-domain sequential recommendation,'' in \emph{Proceedings of the 30th ACM International Conference on Multimedia}, 2022, pp. 471--481.

\bibitem{guo2021gcn}
L.~Guo, L.~Tang, T.~Chen, L.~Zhu, Q.~V.~H. Nguyen, and H.~Yin, ``Da-gcn: A domain-aware attentive graph convolution network for shared-account cross-domain sequential recommendation,'' \emph{arXiv preprint arXiv:2105.03300}, 2021.

\bibitem{li2021dual}
P.~Li, Z.~Jiang, M.~Que, Y.~Hu, and A.~Tuzhilin, ``Dual attentive sequential learning for cross-domain click-through rate prediction,'' in \emph{Proceedings of the 27th ACM SIGKDD conference on knowledge discovery \& data mining}, 2021, pp. 3172--3180.

\bibitem{vaswani2017attention}
A.~Vaswani, N.~Shazeer, N.~Parmar, J.~Uszkoreit, L.~Jones, A.~N. Gomez, {\L}.~Kaiser, and I.~Polosukhin, ``Attention is all you need,'' in \emph{Advances in neural information processing systems}, 2017, pp. 5998--6008.

\bibitem{gray2011entropy}
R.~M. Gray, \emph{Entropy and information theory}.\hskip 1em plus 0.5em minus 0.4em\relax Springer Science \& Business Media, 2011.

\bibitem{chen2022intent}
Y.~Chen, Z.~Liu, J.~Li, J.~McAuley, and C.~Xiong, ``Intent contrastive learning for sequential recommendation,'' in \emph{Proceedings of the ACM Web Conference 2022}, 2022, pp. 2172--2182.

\bibitem{fan2022sequential}
Z.~Fan, Z.~Liu, Y.~Wang, A.~Wang, Z.~Nazari, L.~Zheng, H.~Peng, and P.~S. Yu, ``Sequential recommendation via stochastic self-attention,'' in \emph{Proceedings of the ACM Web Conference 2022}, 2022, pp. 2036--2047.

\bibitem{yuan2022tenrec}
G.~Yuan, F.~Yuan, Y.~Li, B.~Kong, S.~Li, L.~Chen, M.~Yang, C.~Yu, B.~Hu, Z.~Li \emph{et~al.}, ``Tenrec: A large-scale multipurpose benchmark dataset for recommender systems,'' \emph{Advances in Neural Information Processing Systems}, vol.~35, pp. 11\,480--11\,493, 2022.

\bibitem{gao2022kuairec}
C.~Gao, S.~Li, W.~Lei, J.~Chen, B.~Li, P.~Jiang, X.~He, J.~Mao, and T.-S. Chua, ``Kuairec: A fully-observed dataset and insights for evaluating recommender systems,'' in \emph{Proceedings of the 31st ACM International Conference on Information \& Knowledge Management}, 2022, pp. 540--550.

\end{thebibliography}
}

%








\end{document}